%% file: draft.tex
\documentclass[aps,prd,reprint,secnumarabic,nonbibnotes,floatfix,amssymb,longbibliography,showpacs,preprintnumbers]{revtex4-1}

\usepackage{graphicx}
\usepackage{subfigure}
\usepackage{mathptmx}
\usepackage{helvet}
\usepackage{lineno}
\usepackage{textcomp}
\usepackage{multirow}
\usepackage{amsmath}
\usepackage[hyperindex,breaklinks]{hyperref} 

\begin{document}

\preprint{CERN-PH-EP-2011-197} \preprint{Submitted to Physical Review D} 

\title{ Measurement of the azimuthal ordering of charged hadrons with the ATLAS detector }

\author{G.~Aad \textit{et al.}  (The ATLAS Collaboration)} 
\email[Full author list given at the end of the article.]{}

\begin{abstract}   
This paper presents a study of the possible ordering of charged hadrons in the 
azimuthal angle relative to the beam axis in high-energy proton-proton collisions at the Large Hadron Collider (LHC). 
A spectral analysis of correlations between longitudinal and transverse components of 
the momentum of the charged hadrons, driven by the search for phenomena related to the structure of the QCD field,  is performed.
Data were recorded with the ATLAS detector at centre-of-mass energies of $\sqrt{s}$ = 900 GeV and $\sqrt{s}$ = 7 TeV. 
The correlations measured in a kinematic region dominated by low-$p_{T}$ particles 
are not well described by conventional models of hadron production.
The measured spectra show features consistent with the fragmentation
of a QCD string represented by a helix-like ordered gluon chain.
\end{abstract}

\pacs{13.85.Hd,13.87.Fh}

\maketitle

\input{introduction.tex}

\input{observables.tex}
\input{atlas_detector.tex}
\input{mc_samples.tex}
\input{selection.tex}
\input{analysis.tex}

\input{correction_results.tex}

\input{conclusions.tex}

\input{acknowledgements.tex}

\input{appendixA.tex}
\input{appendixB.tex}

\bibliography{draft}
\onecolumngrid
\clearpage
\input{atlas_authlist_arXiv}

\end{document}

%% file: introduction.tex
\section{Introduction}

Inclusive charged-particle distributions have been measured in proton-proton (\textit{pp}) 
collisions at the LHC for different centre-of-mass energies 
\cite{ATLAS-MB1, ATLAS-MB2, ALICE-MB1, ALICE-MB2, ALICE-MB3, CMS-MB1, CMS-MB2, CMS-MB3}. 
These measurements provide insight into the strong interaction (QCD) at low energy scales 
and show that predictions of current phenomenological models
cannot fully describe the measured observables in all kinematic regions. 
 As shown in Ref. \cite{ATLAS-MB2}, the limitation of models is particularly apparent 
when particles with low transverse momentum ($p_{T} <$ 500 MeV) are studied. 
Specifically, discrepancies are observed in the description of
\begin{itemize}
\item{ the charged particle density as a function of pseudorapidity, d$N_{ch}$/d$\eta$;}
\item{ the shape of the charged-particle multiplicity distribution both at low and high multiplicities;} 
\item{ the shape of the charged-particle multiplicity distribution for both $p_{T}< 500$ MeV and $p_{T} > 3$ GeV; and}
\item{ correlations between the average $p_{T}$ of charged particles per event, $\langle p_{T} \rangle$, and the charged-particle multiplicity $n_{ch}$.}
\end{itemize} 
Some of these discrepancies may be reduced 
through the development of parametrisations for the models of non-perturbative QCD and their adjustments (tunes) 
that better match model predictions to the 
latest measurements with particles produced at very low $p_T$. 
Nevertheless, it is also possible that a new  formulation of 
certain components of these phenomenological models is needed.

Many of the difficulties in accurately describing observables dominated by low-$p_T$ QCD phenomena 
stem from the fact that there is often a combination of non-perturbative effects, including soft 
diffraction, low-$p_{T}$ parton scattering and hadronisation. These effects act simultaneously 
in a given kinematic region and are difficult to separate experimentally. 
The recent ATLAS study of inclusive charged-particle production~\cite{ATLAS-MB2} shows 
that the sensitivity of measurements to some of these effects depends on the definition of 
the kinematic region in which the measurements are made. Typically, 
the more inclusively the sample is defined, or the more soft particle activity
is added to the measurement, the larger the disagreement is between the data and the models. 

Hadronisation, the mechanism of hadron formation from quark and gluon 
cascades, must be taken into account in all 
measurements involving hadrons. The flows of energy, momentum and flavour 
of hadrons approximately follow those of partons~\cite{duality}. The mechanism of 
hadron formation, however, can only be described with the aid of phenomenological models. 
The two main models employed in multi-purpose event generators are the string (Lund) 
fragmentation model~\cite{lund} and the cluster model~\cite{cluster01}.

A proposed modification of the Lund string model~\cite{lund_helixm} evokes the possibility of the 
formation of a helix-like structured gluon field at the end of the parton cascade.
Such a structure corresponds to the optimal packing of soft gluons in 
the phase space under helicity conservation requirements. Most notably, the helix string 
model imposes correlations between the adjacent break-up points along a string 
with observable effects in the inclusive $p_{T}$ distribution and in the azimuthal 
ordering of direct hadrons, i.e. hadrons produced directly from the string fragmentation.

In this paper, the correlations in the charged hadron azimuthal angle $\phi$ 
\footnote{The ATLAS reference system is a right-handed coordinate system with its origin at the nominal
 interaction point at the centre of the detector. Cylindrical coordinates ($r,\phi$) are used in the transverse plane,
 $\phi$ being the azimuthal angle around the beam axis. The pseudorapidity is defined in terms of the polar
  angle $\theta$ as $\eta = -\ln \tan~( \theta/2 )$.} 
are studied for two different schemes of hadron ordering using data recorded with the ATLAS detector. 
The results are corrected for detector 
effects and compared with the predictions 
of various Monte Carlo (MC) event generators.
The paper is organised as follows: the observables are described in Section II. 
Section III describes the ATLAS detector and Section IV presents the MC samples used in the study. 
The track and event selection are detailed in Section V. The analysis and the uncorrected data 
distributions measured at the center-of-mass energy of $\sqrt{s}$=7 TeV are presented in Section VI. The procedure to correct for detector and selection 
effects and an overview of systematic uncertainties can be found in Section VII (additional information is provided in the Appendix B). 
The last section of the paper is devoted to a comparison of corrected data with models 
and a discussion of the results.  Appendix A presents results with $\sqrt{s}$=900 GeV.

%% file: observables.tex
\section {Observable quantity}
 
The presence of azimuthal ordering, stemming from the 
underlying QCD string structure, is studied for two different variants of 
the helix-like ordered gluon field. As suggested in \cite{lund_helixm}, 
the azimuthal ordering of hadrons should be observable with the help of a 
power spectrum defined according to the expected structure of the helix field. 
Assuming the break-up of the string occurs via tunneling \cite{lund}, with 
the partons emerging at rest, the azimuthal direction of the hadron's intrinsic 
transverse momentum coincides with the phase of the helix string in the centre 
of the string piece which forms the hadron. Hence, the azimuthal opening angle of two direct hadrons 
measures the phase difference between two corresponding points along the 
string, with the transverse plane defined with respect to the string axis. 
The analysis takes advantage of the fact that in soft \textit{pp} interactions, 
the QCD strings tend to be aligned along the 
beam axis. 

In close analogy with \cite{lund_helixm}, where the authors assume the helix 
winding is proportional to the rapidity difference between hadrons, we define the power spectrum 
  \begin{equation}
     S_{\eta}(\xi) =  \frac{1}{N_{\rm ev}} \sum_{\rm event} \frac{1}{n_{\rm ch}} |\sum_{\rm j}^{n_{\rm ch}}\exp(i(\xi \eta_{\rm j} - \phi_{\rm j}))|^2 ,
 \end{equation} 
where $\xi$ is a parameter and $\eta_{\rm j}$ ($\phi_{\rm j}$) is the pseudorapidity 
(azimuthal angle) of the j-th hadron, $N_{\rm ev}$ is the number of events, and $n_{\rm ch}$ 
is the number of charged hadrons in the event. The inner sum runs over charged 
hadrons in the event and the outer sum over events in the sample.

It is important to note that the form of the helix field is not well constrained 
and that it is possible to find several parametrisations of the helix field conforming 
to the assumptions made in  \cite{lund_helixm}. One possible scenario \cite{ismd} 
corresponds to a static, regular helix structure with the phase difference $\Delta\phi$ 
proportional to the amount of internal energy stored in the fraction $f$ of string with mass $M_0$ 
 \begin{equation}
   \Delta\phi = \mathcal{S} \ f \ M_{0} = \mathcal{S} \ \kappa \ \Delta l  = \mathcal{S} \ \Delta E  ,
\end{equation} 
where $\mathcal{S} $ is a parameter, $\kappa$ is the string energy density, 
$\Delta\phi$ is the difference in the helix phase between two points along 
the string and $\Delta l$  and $\Delta E$  are the length and the energy, 
respectively, of the corresponding string piece in its rest frame. 
The energy-distance $\Delta E$ along the string between direct hadrons is 
not directly observable but according to MC studies, the signature of the helix-shaped field 
should be visible with help of a very loose approximation of the 
string by a chain of hadrons ordered in pseudorapidity.
 For the purpose of measuring the azimuthal ordering, we 
thus retain two parameters for each final hadron: the azimuthal angle $\phi$
 and the position $X$ along the chain, evaluated as
\begin{equation}
\label{eq:chain}
      X_{\rm j}   =  0.5 \ E_{\rm j} + \sum_{\rm k=0}^{\rm k<j} E_{\rm k} ,
\end{equation} 
where $E_{\rm k}$ is the energy of the k-th hadron in the chain, and the position of 
the hadron is associated with the centre of the corresponding string piece. Accordingly, we 
define an alternative power spectrum
\begin{equation}
     S_{E}(\omega) =  \frac{1}{N_{\rm ev}} \sum_{\rm event} \frac{1}{n_{\rm ch}} |\sum_{\rm j}^{n_{\rm ch}}\exp(i(\omega X_{\rm j} - \phi_{\rm j}))|^2 ,
\end{equation}
where $\omega$ is a parameter. The inner sum runs over pseudorapidity-ordered 
charged hadrons in the event.

The presence of a helix-like angular ordering of hadrons of either type would 
manifest itself as a peak in the corresponding power spectrum; the position of the peak would
indicate the density of the helix winding. It should be 
stressed that, though formally very similar, $S_{E}$ and $S_{\eta}$ are only 
loosely correlated. A modified form of the helix string implies a difference in the 
experimental signature, such that the presence of a helix gluon field creating a peak in
$S_{E}$ does not necessarily result in a peak structure in $S_{\eta}$ and vice-versa.

   The power spectra can also be expressed as a sum of contributions from pairs of hadrons
\begin{eqnarray}
     S_{\eta}(\xi) & = & 1 \ + \frac{1}{N_{\rm ev}} \sum_{\rm event} \frac{1}{n_{\rm ch}} \sum_{\rm i\neq j} \text{cos}(\xi \Delta\eta_{\rm ij}- \Delta\phi_{\rm ij}),  \nonumber  \\
     S_{E}(\omega) & = & 1 \ + \frac{1}{N_{\rm ev}} \sum_{\rm event} \frac{1}{n_{\rm ch}} \sum_{\rm i\neq j} \text{cos}(\omega \Delta X_{\rm ij}- \Delta\phi_{\rm ij}),   
\end{eqnarray} 
   where $\Delta\phi_{\rm ij}=\phi_{\rm i}-\phi_{\rm j}$ is the opening azimuthal angle between hadrons, 
$\Delta\eta_{\rm ij}=\eta_{\rm i}-\eta_{\rm j}$ is their pseudorapidity difference and $\Delta X_{\rm ij}=X_{\rm i}-X_{\rm j}$ 
their energy-distance as defined above. The absence of correlations corresponds to $S_{\eta}(\xi)$=1 and
$S_{E}(\omega)$=1. 

%% file: atlas_detector.tex
\section{ATLAS detector}

The ATLAS detector \cite{ATLAS} covers almost the entire solid angle around the 
collision point with layers of tracking detectors, calorimeters and muon chambers. 
It has been designed to study a wide range of physics topics at LHC energies.  
For the measurements presented in this paper, the trigger system and the tracking devices 
are of particular importance.

The ATLAS inner detector has full coverage in $\phi$ and covers the pseudorapidity 
range $|\eta| < 2.5$. It consists of a silicon pixel detector, a silicon strip detector 
(SCT) and a transition radiation tracker (TRT). These detectors are immersed in a 2~T axial magnetic field. The pixel, SCT and TRT detectors have typical  $r$--$\phi$ position resolutions of 10, 17 and 130 $\mu$m, respectively,  and  the pixel and SCT detectors have $r$--$z$ position resolutions of 115 and 580 $\mu$m, respectively. 
A track traversing the full radial extent would typically have 3 silicon pixel hits, 8 and more silicon strip hits  and more than 30 TRT hits.

The ATLAS detector has a three-level trigger system:~level~1~(L1), level~2 (L2) and the 
event~filter~(EF).
For this measurement, the L1 trigger relies on the beam pickup timing devices (BPTX) and
the minimum bias trigger scintillators (MBTS).
The BPTX are composed of electrostatic beam pickups attached to the beam pipe at a 
distance $z = \pm 175$~m from the center of the ATLAS detector.
The MBTS are mounted at each end of the detector in front of the liquid-argon endcap-calorimeter 
cryostats at $z = \pm 3.56$~m and are segmented into eight sectors in azimuth 
and two rings in pseudorapidity ($2.09 < |\eta| < 2.82$ and $2.82 < |\eta| < 3.84$).
Data were taken for this analysis using the single-arm MBTS trigger, formed from BPTX 
and MBTS L1 trigger signals.
The MBTS trigger was configured to require one hit above threshold from either side of the 
detector. The MBTS trigger efficiency was studied with a separate pre-scaled L1 BPTX trigger, 
filtered to obtain inelastic interactions by inner detector requirements at L2 and EF \cite{ATLAS-MB2}.

%% file: mc_samples.tex
\section{Monte Carlo samples }

The analysis has been developed using a PYTHIA 6.421~\cite{PYTHIA} 
(MC09 tune~\cite{ATLAS-MC09}) sample with large statistics, 
fully simulated~\cite{GEANT4} and reconstructed using the standard ATLAS 
reconstruction chain~\cite{ATLAS-simul}. It should be stressed that the choice of the
MC tune does not affect the outcome of the analysis because the data are corrected in a model independent way.  

For the comparison of the corrected data with the standard hadronisation models, 
MC samples produced with PHOJET 1.12.1.35~\cite{PHOJET}, 
HERWIG++ 2.5.1~\cite{HERWIG++} (LHC-UE7-2/MU900-2 tunes~\cite{hwgtune}), PYTHIA 8.130 (4C tune~\cite{PYTHIA8}) and a recent tune of PYTHIA 6 (AMBT2b~\cite{AMBT2b}) 
have been used. The MC generators PYTHIA and PHOJET employ the Lund string 
fragmentation model whereas the HERWIG MC is based on the cluster model.
To study the sensitivity of the power spectra to the modification of the string
fragmentation model, the data are also compared with an alternative implementation of the 
fragmentation process based on the helix string field described by Eq. 2~\cite{pystrf}. 

%% file: selection.tex
\section {Data samples}    
 
The measurements reported in this paper were made using \textit{pp} collision 
data recorded at $\sqrt{s}=7$ TeV. The data were collected with stable colliding 
beams at 7 TeV and correspond to an integrated luminosity of $\sim  190~ \mu$b$^{-1} $ 
from the beginning of the 2010 LHC run~\cite{ATLAS-MB2}. A sample of \textit{pp} collision 
events recorded at $\sqrt{s}=900$ GeV corresponding to an integrated luminosity of 
$\sim  7~ \mu$b$^{-1} $ \cite{ATLAS-MB2} 
was also studied and the results are shown in Appendix A.
    
\subsection {Event and track selection} 

Events are selected using the following criteria:
\begin{itemize} 
\item{ the event has at least one trigger hit in MBTS; }
\item{ the event has one and only one reconstructed vertex and this vertex must have at 
least three associated tracks; }
\item{ the event has no tracks with $p_{T} > 10$ GeV;}
\item{ the event has at least six reconstructed tracks ($n_{tr}>5$) passing the requirements below.}
\end{itemize}

The requirements on reconstructed tracks included in the analysis are the following:
\begin{itemize}
\item{ the track is reconstructed by the track reconstruction algorithm used in 
\cite{ATLAS-MB1}, with an implicit cut on the transverse momentum, $p_T>$ 100~MeV, 
and more than 6 hits in the silicon detectors; }
\item{ the track has a transverse impact parameter with respect to the primary vertex  $|d_{0}^{PV}| < $ 2 mm;}
\item{ the track has a longitudinal impact parameter with respect to the primary vertex  $|z_{0}^{PV}| \sin(\theta) <$ 2 mm; and}
\item{ the track is reconstructed in the pseudorapidity range $-2.5 < \eta < 2.5$.}
\end{itemize}

The requirement on the minimum number of tracks ensures full trigger efficiency~\cite{ATLAS-MB2}. 
The contributions from the beam and non-collision background (cosmic rays and detector noise) have been 
investigated in~\cite{ATLAS-MB2} and found to be negligible. 
Events with multiple primary vertices (less than 0.3\% of the sample and subsamples defined below) 
are rejected in order to prevent a bias from multiple $pp$ interactions in the colliding proton bunches.

\subsection{Subsample definitions}

The analysis is carried out in parallel on the sample selected as described above 
(henceforth referred to as the ``inclusive sample'') and on two subsamples.

The first subsample contains events where the transverse momentum of any reconstructed track does 
not exceed 1~GeV (max$(p_{T})<$ 1~GeV). This subsample is called the ``low-$p_{T}$ enhanced sample''. 
The effects of parton showering and lateral boost are diminished in this selection. 
The max$(p_T)< 1$~GeV requirement selects events with little acollinear jet activity and thus the transverse activity is expected to be primarily sensitive to hadronisation effects. 

The analysis is also performed on a second subsample defined by a higher
track $p_T$ cut-off, $p_T>$ 500~MeV. This particular selection yields a subsample with 
a significantly reduced contribution from diffractive \textit{pp} interactions (see Table I). This
subsample is referred to as the ``low-$p_{T}$ depleted sample''.

\subsection{Selection criteria at particle level} 
 
The comparison between corrected data and MC models requires an adjustment of the event selection 
in order to avoid a systematic bias. The analysis relies on two main selection criteria: 
the charged-hadron $p_T$ and the charged-hadron multiplicity.

The effect of the low $p_T$ cut-off is easily modelled at the particle level (MC truth \footnote{Primary particles are defined as all 
particles with lifetime longer than 0.3$\times$10$^{-10}$s originating from the primary interaction 
or from subsequent decay of particles with shorter lifetime.}) 
and the corresponding systematic uncertainty is covered by the uncertainty assigned to the 
correction procedure, described in Section~VII. The cut on the maximal $p_T$ of a 
track is more selective when applied at the particle level as it removes also those MC events 
which contain a non-reconstructed  high $p_T$ track. The effect is non-negligible 
in the low-p$_T$ enhanced sample, where it is corrected for and a systematic uncertainty 
is assigned to reflect the additional uncertainty.

 The charged-track multiplicity selection criteria on the measured data need to be modified in order to emulate
MC modelling. The adjustment of the charged-track multiplicity is done in the following way: for each reconstructed track,
 a random number RND is repeatedly generated according to a flat distribution until $RND<\epsilon_{tr}$, $\epsilon_{tr}$ being the estimated track reconstruction efficiency  \footnote{the track reconstruction efficiency is parametrised as 
a function of track $p_{T}$ and pseudorapidity $\eta$}.  The corrected charged-track mutiplicity $n_{ch}$ corresponds 
 to the number of random numbers generated for the entire event. The procedure also contains an additional correction for 
the residual content of secondary tracks. 

\begin{figure}[b]
\begin{center}
\includegraphics[width=0.45\textwidth]{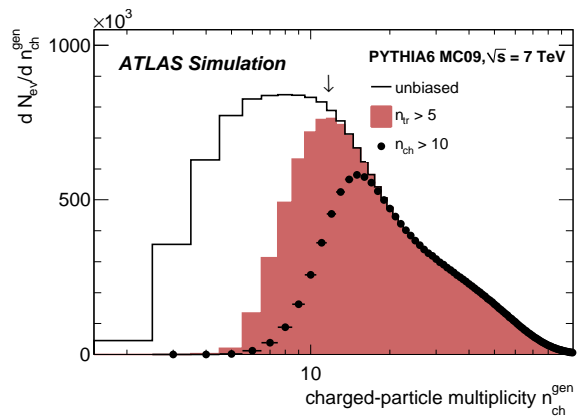}
\caption{
 The impact of the charged-track multiplicity cut-off on the true charged-particle multiplicity distribution.   
The arrow indicates the cut on the true charged-particle multiplicity, removing 
the region affected by the loss of events due to the requirement of at least 6 
reconstructed tracks (white area). The shaded area corresponds to the detector-level cut $n_{tr}>5$.
The final analysis selection cut  ($n_{ch}>10$)  is indicated by closed points.
\label{fig:nch}}
\end{center} 
\end{figure}

The selection based on $n_{ch}$ roughly reproduces the average 
charged-particle multiplicity of the particle-level sample selected with a 
$n_{ch}^{gen}$ cut-off (see Table I). Figure~\ref{fig:nch} illustrates 
the effect of hadron-level cuts on the true charged-particle multiplicity 
$n_{ch}^{gen}$. The choice of the selection cut for the current analysis 
($n_{ch}\approx n_{ch}^{gen}>$ 10 ) is aimed at minimising any bias in the power spectra 
related to the loss of events due to the detector-level charged-track multiplicity 
cut-off $n_{tr}>$5. 

The selection criteria distinguishing between data samples are summarised in Table II which also provides the information about the final number
 of events retained for the analysis and the mean corrected charged-track multiplicity of each sample.
The relative fraction of diffractive events (based on the nominal cross-section) and the mean 
charged-particle multiplicity of the Monte Carlo selection are given in Table I.

The final track selection contains 2.8$\pm$0.4\% of secondary tracks according to studies 
performed on simulated samples. Non-primary tracks predominantly arise from hadronic interactions 
with detector material,
photon conversion to electron-positron pairs and decays of long lived particles.
The average reconstruction efficiency for primary charged particles is 75\% for 
the $p_T>$100 MeV track selection and 84\% for the $p_T>$500 MeV selection. 
The systematic uncertainty due to the performance of the track reconstruction is estimated
following the studies performed in Ref.~\cite{ATLAS-MB2}.
\begin{table}[h!]{
\begin{center}
\begin{tabular}{|l|c|c|c|}
\hline
 Model(tune) &  \multicolumn{3}{|c|}{ PYTHIA6(MC09)}  \\ 
\hline 
low $p_{T}$ cut &   \multicolumn{2}{|c|} {  $>$100 MeV  } &  $>$500 MeV \\
\hline
max($p_{T}$) cut &  $<$ 10 GeV & $<$ 1 GeV & $<$ 10 GeV \\   
\hline
 & \multicolumn{3}{|c|}{  n$_{ch}>$10  (corrected detector level)  } \\
\hline
$N^{gen}_{ch}$   &  31.93  &  17.11$^*$ & 22.62  \\
\hline
diffractive/total &  3.9$\pm$0.1\%  &  21.4$\pm$0.2\%  & $<$0.1\%  \\  
\hline
 & \multicolumn{3}{|c|}{  n$^{gen}_{ch}>$10  (particle level)  } \\
\hline
$N^{gen}_{ch}$   &  31.31  &  15.53 & 22.26  \\
\hline
diffractive/total &   4.0$\pm$0.1\%  &  22.7$\pm$0.2\%  & $<$0.1\%  \\  
\hline
\end{tabular}
\label{tab:sel_mc}
\caption{Average charged-particle multiplicity $N^{gen}_{ch}$ and relative fraction of diffractive events
for the fully simulated PYTHIA6 (MC09) MC sample  at $\sqrt{s}=7$ TeV. 
Results are shown for events selected with the corrected charged-track multiplicity 
cut-off $n_{ch}>$10 (detector level) and  $n_{ch}^{gen}>$10 (particle level) 
($^*$ indicates before the max($p_T$) cut correction). }
\end{center}}
\end{table}      
\begin{table}[h!]{
\begin{center}
\begin{tabular}{|c|c|c|c|c|c|}
\hline
\multicolumn{6}{|c|} { \textit{pp} collisions at $\sqrt{s}$ = 7 TeV, \textit{n}$_{ch}>$10 } \\
\hline 
\multicolumn{4}{|c|} {  $p_T>$100 MeV  } &  \multicolumn{2}{|c|} { $p_T>$500 MeV }  \\
\hline 
\multicolumn{2}{|c|} {  max($p_T$)$<$10 GeV } &  \multicolumn{2}{|c|} { max($p_T$)$<$1 GeV } &   
\multicolumn{2}{|c|} { max($p_T$)$<$10 GeV  } \\
\hline
$N_{ev}$   &  $ N_{ch} $   &  $N_{ev}$   &  $ N_{ch} $  & $N_{ev}$   &  $ N_{ch} $  \\
\hline
8,099,211 & 34.71 &   1,292,389  &  17.96$^*$ &  4,341,217 & 23.27    \\        
\hline
\end{tabular}
\label{tab:sel_data}
\caption{Number of selected data events and average corrected charged-track multiplicity, per sample 
 ($^*$ indicates before the max($p_T$) cut correction).}
\end{center}}
\end{table}      

%% file: analysis.tex
\section{Analysis method }

For the measurement of $S_{E}$, the selected tracks are ordered 
by pseudorapidity and a pion mass is assigned to each of them. According to MC 
estimates, the charged particle sample contains about 86\% pions, 9.5\% kaons, 4\% 
protons/antiprotons and a negligible number of leptons ($\sim$ 0.5\%). The effect
of assuming a pion mass would need to be taken into account for a precision measurement 
of the position of the signal but its impact on the comparison of data with MC models is negligible.
For the calculation of $S_{\eta}$ no mass assumption is required. The power spectra are measured 
as ($S_\eta~ - ~ 1$) and ($S_E~ - ~ 1$) for convenience.
\begin{figure}[t]
\begin{center}
\includegraphics[width=0.46\textwidth]{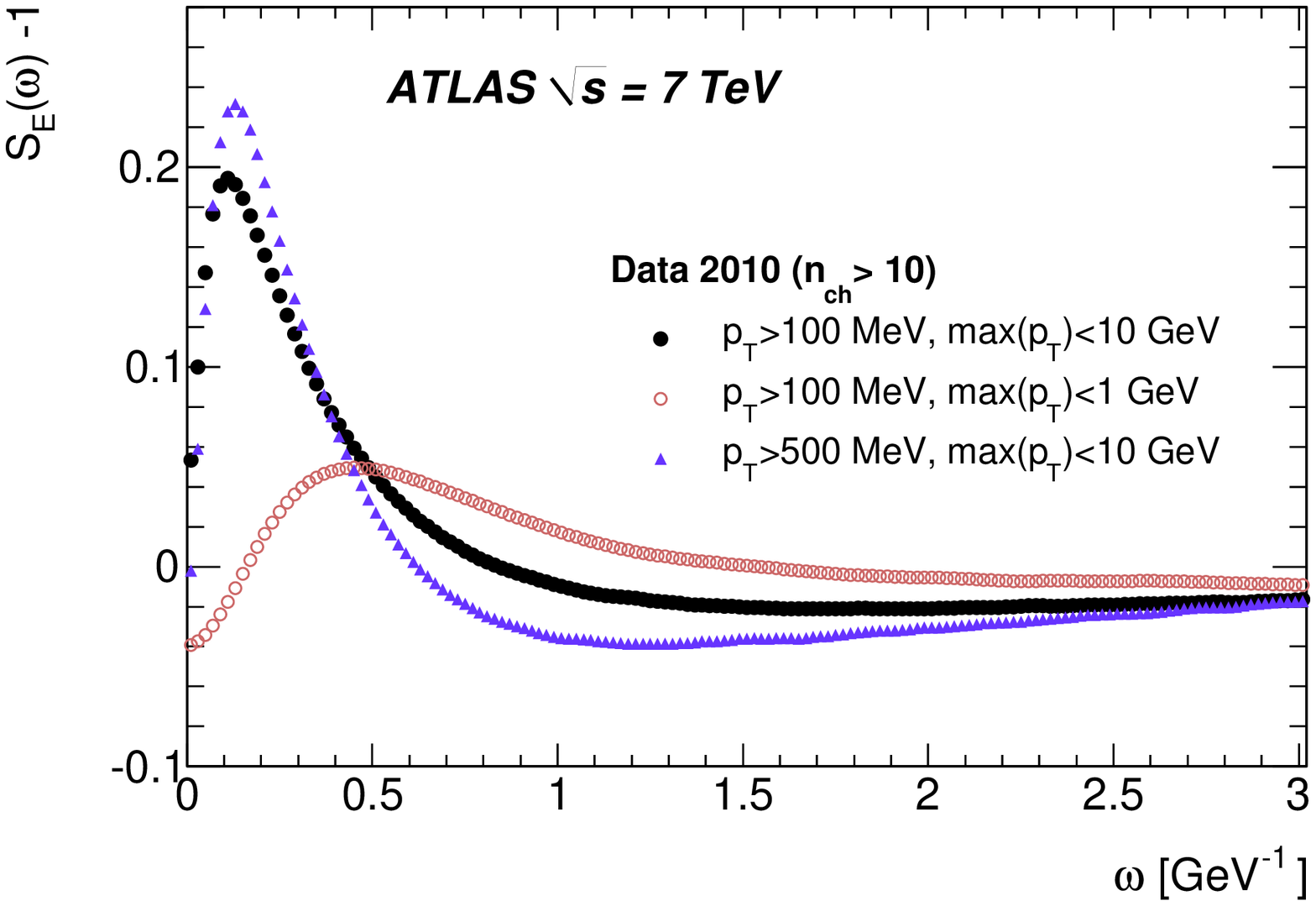}
\includegraphics[width=0.46\textwidth]{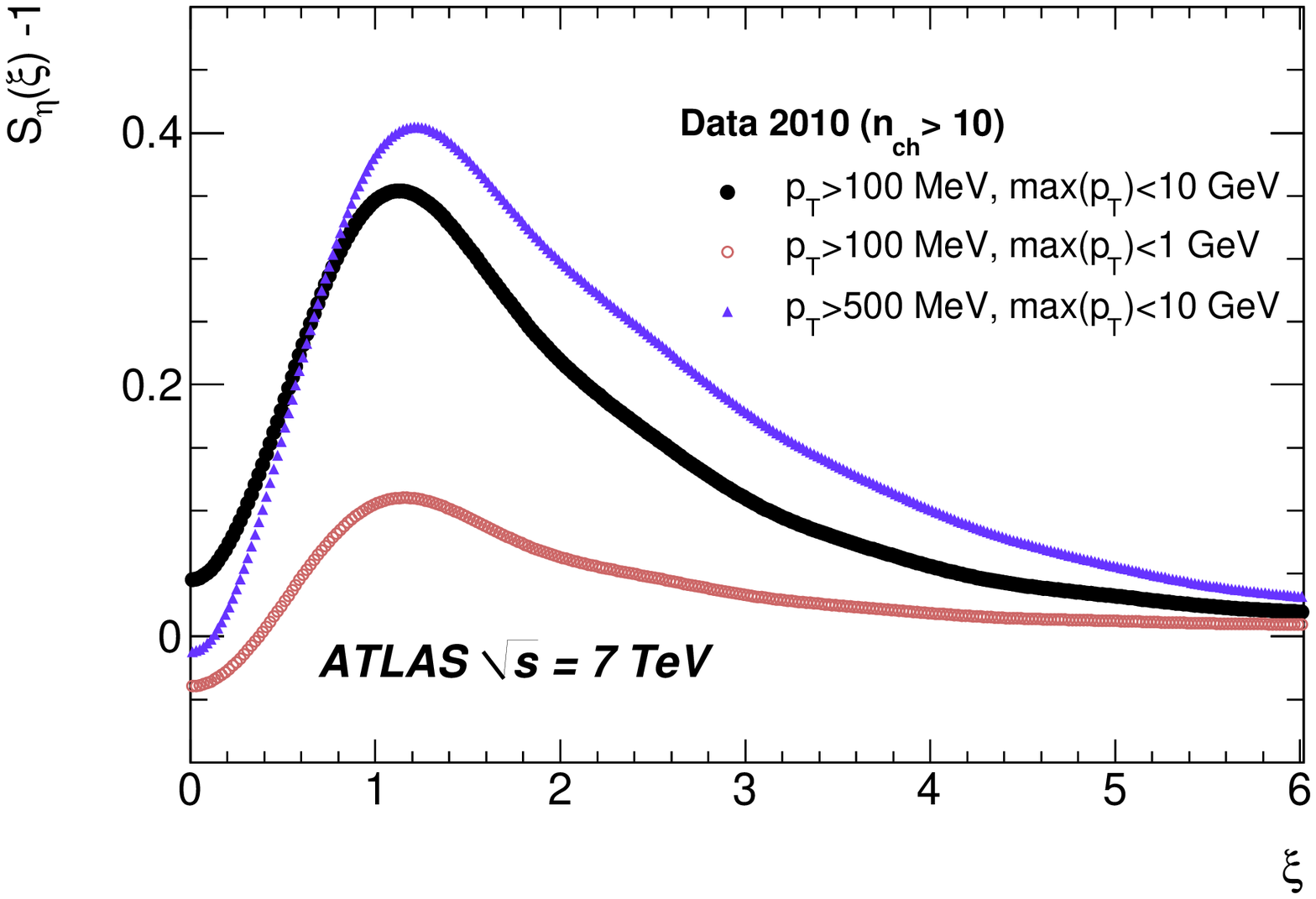}
 \caption{
 The uncorrected data measurements (top: $S_E$, bottom: $S_{\eta}$) obtained from the
 data sample collected at $\sqrt{s}$ = 7 TeV. The measurement of the inclusive sample 
 is compared to measurements of the low-$p_{T}$ enhanced and low-$p_{T}$ depleted 
 subsamples.
 \label{fig:uncorrected}}
\end{center}
\end{figure}

The uncorrected power spectra are shown in Fig.~\ref{fig:uncorrected}
as a function of the azimuthal opening angle (helix phase difference) per unit of energy distance ($S_{E}$)
and per unit of pseudorapidity ($S_{\eta}$) for the three samples. All angles are expressed in radians throughout the paper.  

MC studies show that the power spectra are sensitive to various kinds of
correlations between particles. The dominant peaks seen in both distributions 
arise due to the jet structure and momentum conservation in the hard parton-parton scattering.  
The position of these peaks depends on the visible energy (in the case of $S_{E}$) and the 
pseudorapidity range (in the case of $S_{\eta}$) used in the analysis. Their height is sensitive 
to a number of physics processes, notably the amount and structure of multiple 
parton interactions, cross-talk between overlapping hadronic systems (colour 
reconnection), and parton shower properties. Note, that the values of $S_{E}(\omega=0)$ and 
$S_{\eta}(\xi=0)$ are identical by definition. They are closely related 
to the average opening angle between particles in the transverse plane. 
The presence of azimuthal correlations 
stemming from the properties of the gluon field should be visible as an additional positive 
peak or enhancement in the power spectrum and the modification should be more pronounced
 in the low-p$_{T}$ region, where the fragmentation and parton interactions have comparable 
effects on the transverse momentum of hadrons. 

The comparison of the uncorrected data obtained in the inclusive event selection and in the 
low-$p_{T}$ enhanced/depleted subsamples shows the size of the peaks diminishing with
the decreasing track $p_T$ selection range, a feature we may associate with the relative fraction of 
high $p_T$ jets in the sample.

%% file: correction_results.tex
\section{Correction procedure and systematic uncertainties}

The data are corrected for non-reconstructed charged particles with the help of an unfolding technique 
based on \cite{unfolding} and described in the Appendix B. 
    
The correction for the secondary track content is obtained using a random 
sampling of secondary tracks according to the parametrised secondary track rate 
obtained from fully simulated MC. The contribution from tracks labeled as secondary 
($\delta S^{sec}$) is subtracted from the measured data distribution. Typically, it amounts to $\sim$ 6\% of the
size of the peak in the power spectrum.    
 
An additional correction is applied in the low-$p_T$ enhanced sample to compensate for the 
bias introduced by the selection cut on max($p_T$) in case the track with highest p$_T$ was not reconstructed. Selecting events with exactly one track 
above the threshold, the power spectra ($S^{p_T}$) are calculated using all the other tracks in the event. 
The correction is obtained by subtracting $S^{p_T}$ from the data in the proportion corresponding 
to the probability for the high-$p_T$ track being lost in the reconstruction. 

All corrections are model independent. The correction procedure has been verified by checking the
procedure on fully simulated MC samples. 
 
\subsection{ Systematic uncertainties } 
\begin{table}[b]{
\begin{center}
\begin{tabular}{|c|l|}
\hline
   Source  &   Systematic uncertainty [$S_{E}(\omega)$, $S_{\eta}(\xi)$] \\
\hline \hline
  folding procedure   &  \multicolumn{1}{|c|} { max(0.003, 3\% $(S-1))$ }     \\
\hline
  unfolding   &  \multicolumn{1}{|c|}{ envelope of the residual bias from unfolding }     \\
\hline
  tracking efficiency &  \multicolumn{1}{|c|}{ scaling parameters $\pm$ 5\%}  \\        
\hline
  secondary tracks   &  \multicolumn{1}{|c|} {  max(0.005, 0.25 $|\delta S^{sec}|$) }   \\        
\hline
  $n_{ch}$ cut-off   &  \multicolumn{1}{|c|}{variation of the cut-off $n_{ch} ~\text{by}~ \pm$ 1}    \\        
\hline
  max($p_T)<1$ GeV    &  \multicolumn{1}{|c|}{ $\delta S^{p_T} = (1-\epsilon_{tr})/\epsilon_{tr} \ S^{p_T}, \ \epsilon_{tr}\pm 5\%$ }   \\        
\hline
\end{tabular}
\label{tab:syst_bin}
\caption{The parametrisation of the components of the systematic uncertainty, per measured point of the corrected power spectrum.}
\end{center}}
\end{table}      
\begin{figure}[b!]
\begin{center}
\includegraphics[width=0.48\textwidth]{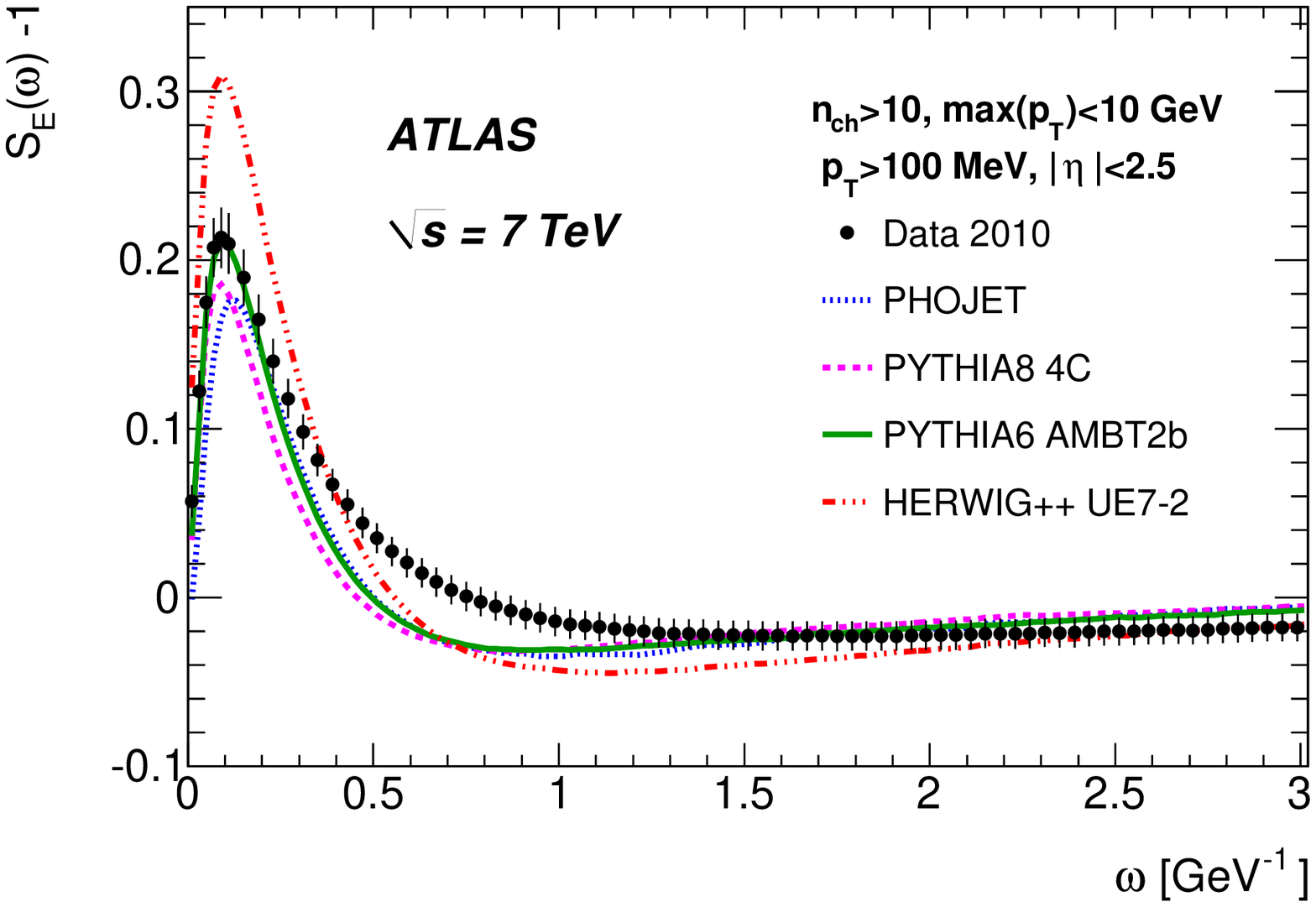}
\includegraphics[width=0.48\textwidth]{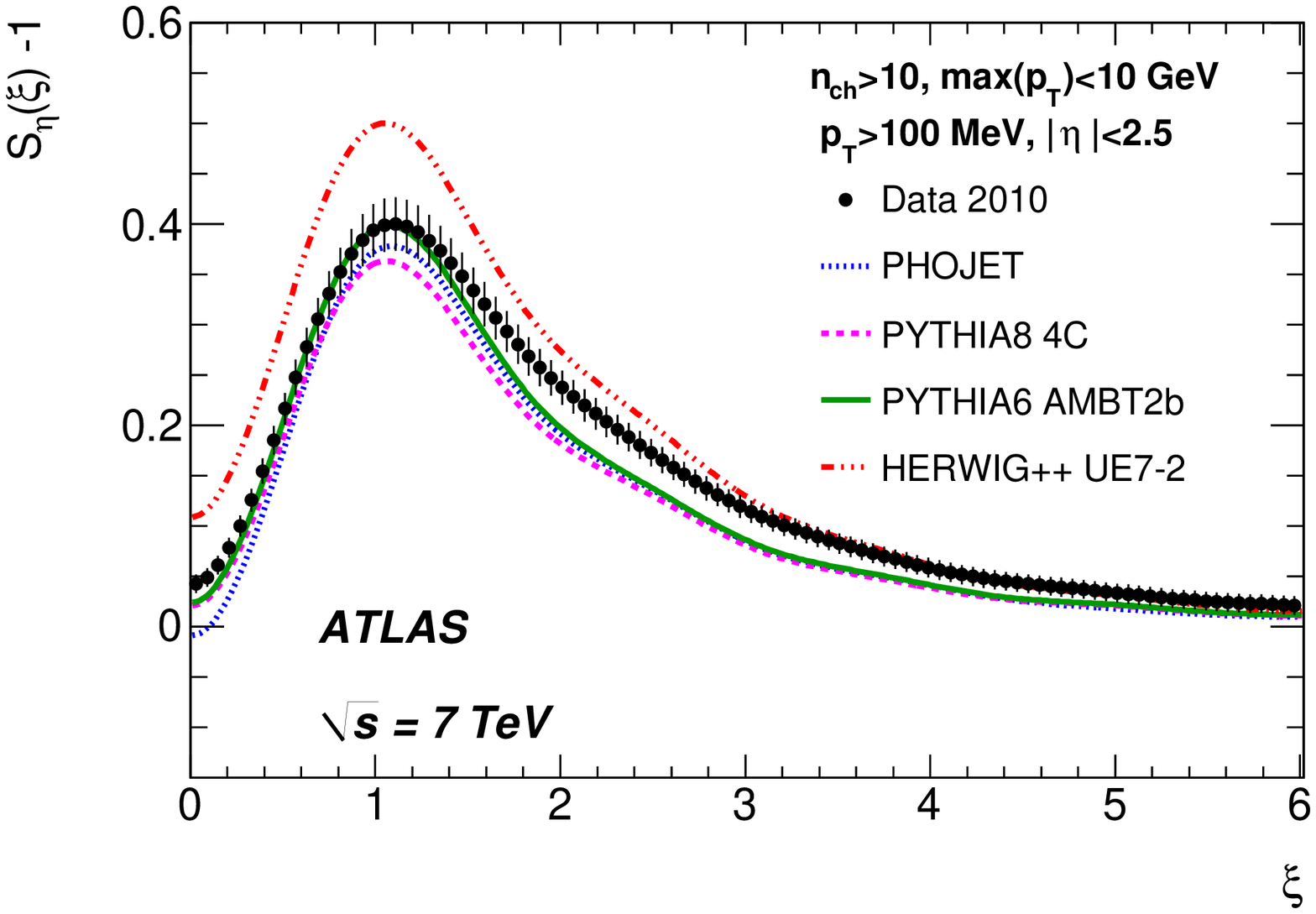}
 \caption{
   Corrected data from the inclusive sample compared to particle-level predictions from various 
   MC models using conventional hadronisation algorithms. The top and bottom plots are for the 
$S_{E}$ and $S_{\eta}$ power spectra, respectively. The error bars correspond to the 
combined statistical and systematic uncertainties.
\label{fig:corr_pt100}}
\end{center}
\end{figure}
The principal sources and parametrisations of systematic uncertainty associated 
with the corrected data are summarised in Table III.
The combined systematic uncertainty 
has the following components:  
\begin{itemize}
\item{residual bias of the folding procedure (Appendix B):} obtained from the comparison of distributions 
reconstructed at the detector level using samples with full detector simulation 
and those obtained with the folding technique; 
\item{uncertainty of the unfolding technique:} parametrised to cover the residual 
discrepancies in the scaling of 3 folding iterations (Appendix B);
\item{uncertainty on the tracking efficiency estimate:} dominated by the uncertainty 
on the inner detector material description, which translates into a variation of scaling factors by 5\%;
\item{uncertainty due to the residual content of secondary tracks:} set to 25\% of the 
correction applied, with minimal value of 0.005 (based on MC studies) ; 
\item{uncertainty due to the difference in the charged-particle multiplicity selection at the generator level 
and at the detector level}: calculated in a model-independent way as a variation of the 
shape corresponding to the change of the averaged selected charged-particle multiplicity by one unit; and
\item{the uncertainty in the correction of the bias due the max($p_T$) cut}: corresponds to a 5\% variation of the track reconstruction efficiency.
\end{itemize}
 All contributions to the systematic uncertainty are combined quadratically. The negative 
correlation between track reconstruction efficiency and secondary track content is neglected, 
making the uncertainty estimate more conservative.

\section{Results}
The results of this analysis obtained for $pp$ collisions at $\sqrt{s}$ = 7 TeV are 
presented in this section.
Results from this analysis repeated for $\sqrt{s}$ = 900 GeV are shown in Appendix A.
The corrected data are compared with the predictions of several commonly used
MC models: PYTHIA6, PHOJET, PYTHIA8 and HERWIG++. 

\begin{figure}[b!]
\begin{center}
\includegraphics[width=0.48\textwidth]{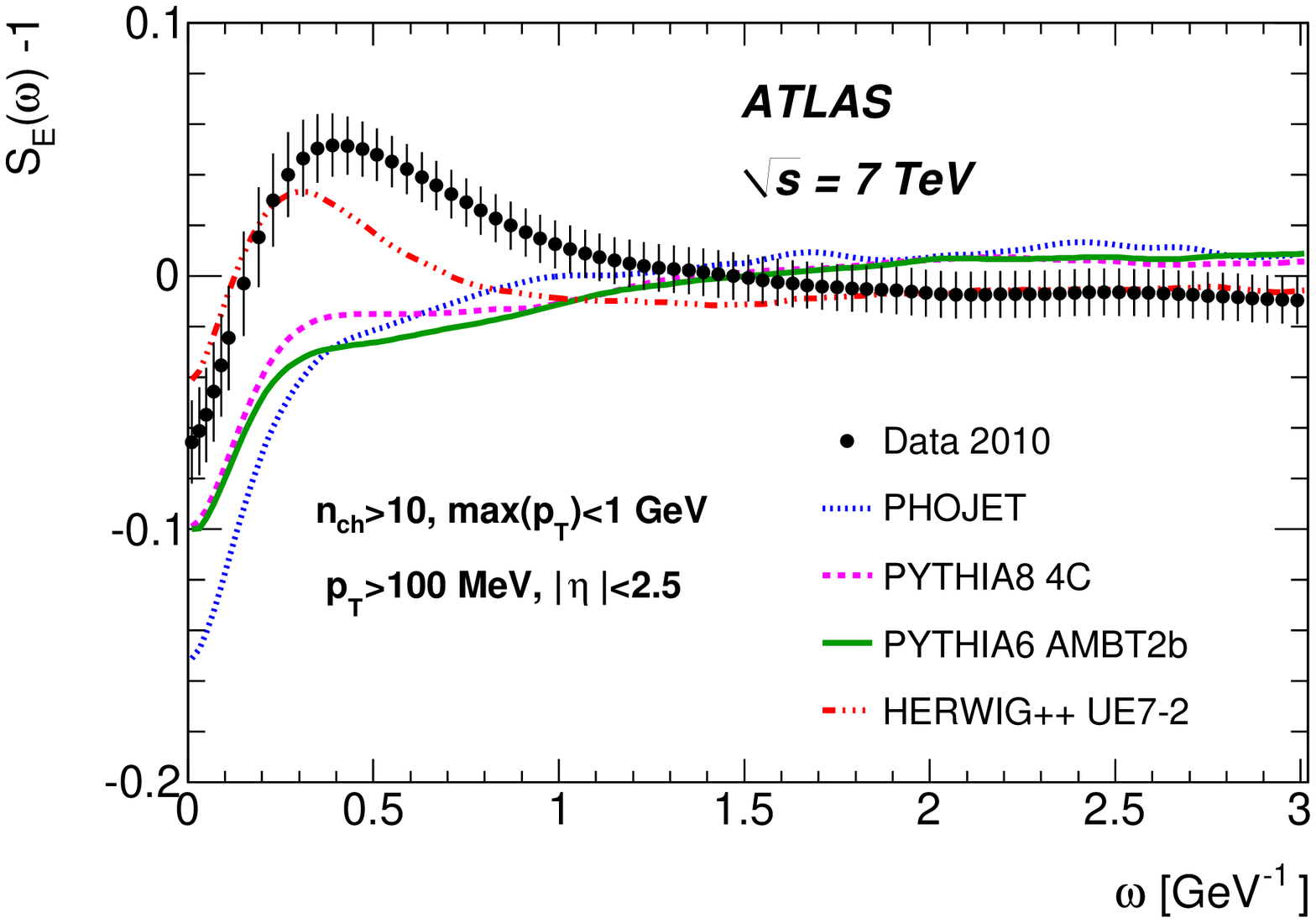}
\includegraphics[width=0.48\textwidth]{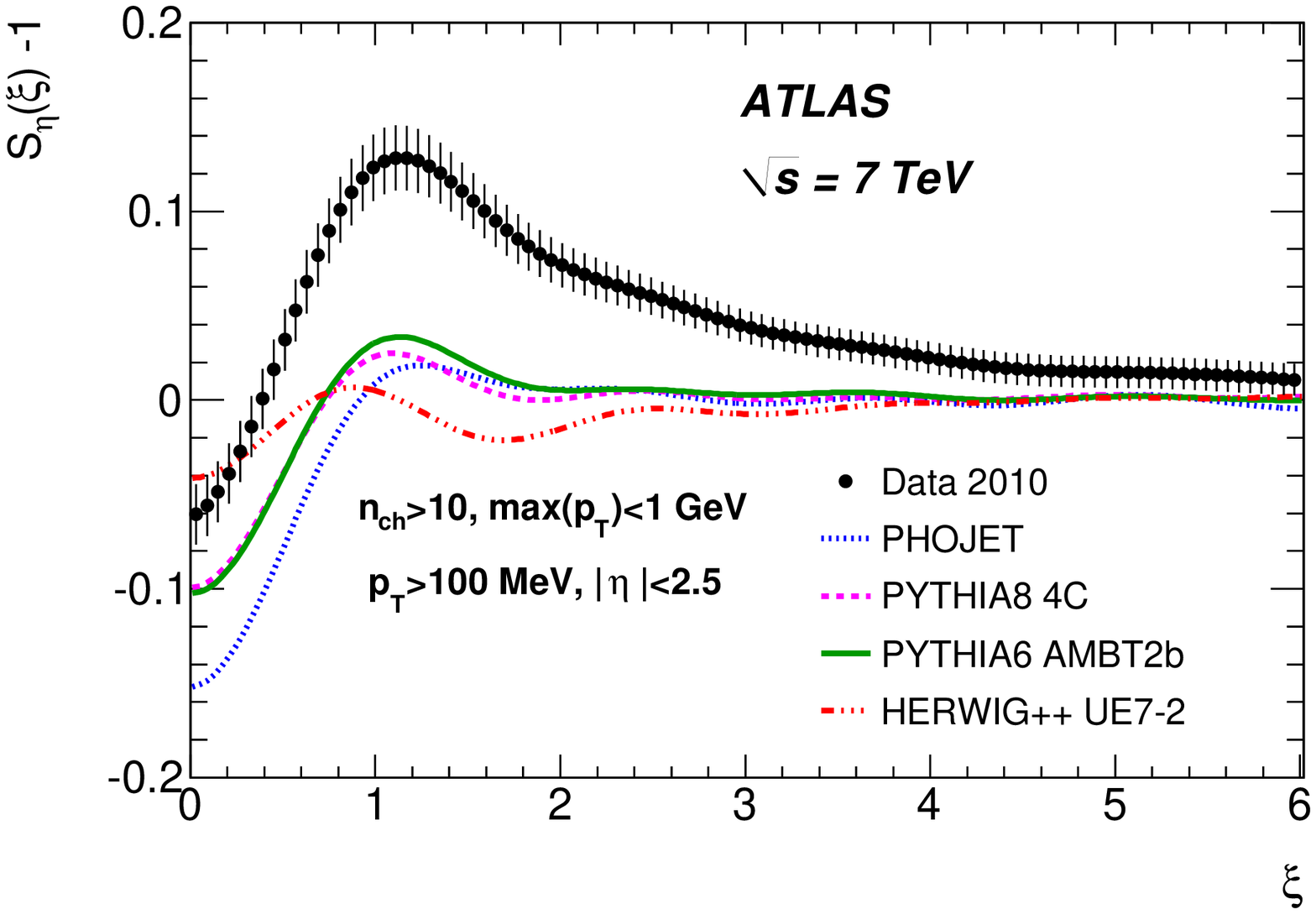}
\caption{
   Corrected data from the low-p$_T$ enhanced sample compared to particle-level predictions from various 
   MC models using conventional hadronisation algorithms. The top and bottom plots are for the 
$S_{E}$ and $S_{\eta}$ power spectra, respectively. The error bars correspond to the 
combined statistical and systematic uncertainties.
\label{fig:corr_pt1}}
\end{center}
\end{figure}

Figure~\ref{fig:corr_pt100} shows the comparison for the inclusive event selection 
($n_{ch}>10$, $p_T>$100 MeV and maximal $p_{T} <$ 10 GeV).
The principal peak structure observed in the power spectra for both $S_{E}$ and $S_{\eta}$ is 
roughly reproduced by PYTHIA and PHOJET models and overestimated by HERWIG++. 
The tail of the $S_E$ distribution around $ 0.5 < \omega < 1$ rad/GeV is not reproduced by any of the models. 

Hadronisation effects should become more evident when 
measurements are made in regions of the 
phase space dominated by the production of low-$p_{T}$ particles.
Figure \ref{fig:corr_pt1} shows the power spectra measured in the low-$p_{T}$ enhanced sample ($n_{ch}>10$, 
$p_T>$100 MeV and maximal $p_{T} <$ 1 GeV).
A significant amount of correlations in observed in the data in both $S_{E}$ and $S_\eta$ distributions
compared to the PHOJET and PYTHIA based models. HERWIG++ gives a seemingly better description for the $S_E$ distribution
yet it seems its prediction is more of an artefact of an enhanced single jet structure given the fact the model
 overestimates the measurements in the inclusive event selection (Fig.~\ref{fig:corr_pt100}).
The interpretation of this measurement in terms of the azimuthal ordering of hadrons related to the properties
of the gluon field is discussed in Section VIII.1.

\begin{figure}[b!]
\begin{center}
\includegraphics[width=0.48\textwidth]{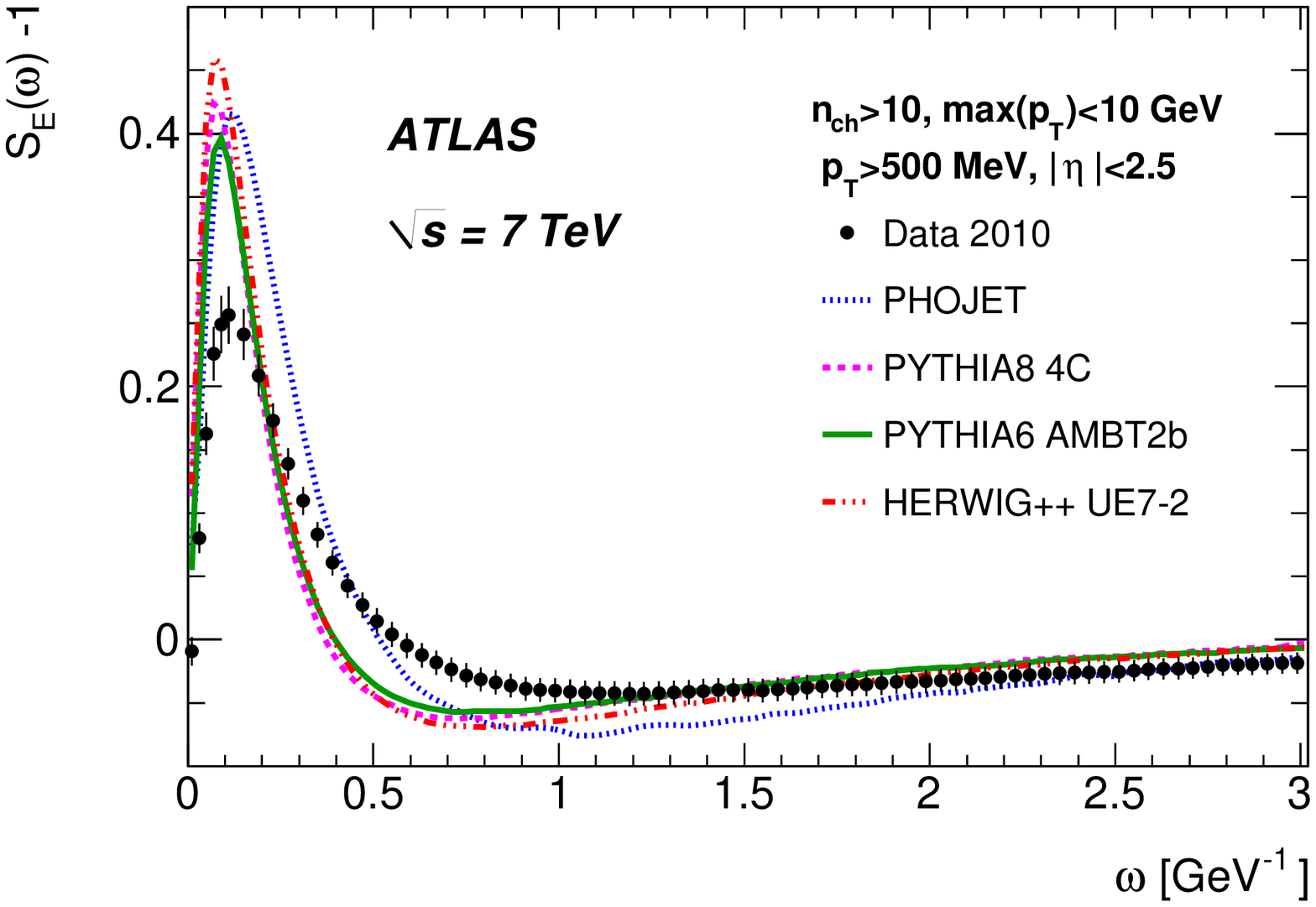}
\includegraphics[width=0.48\textwidth]{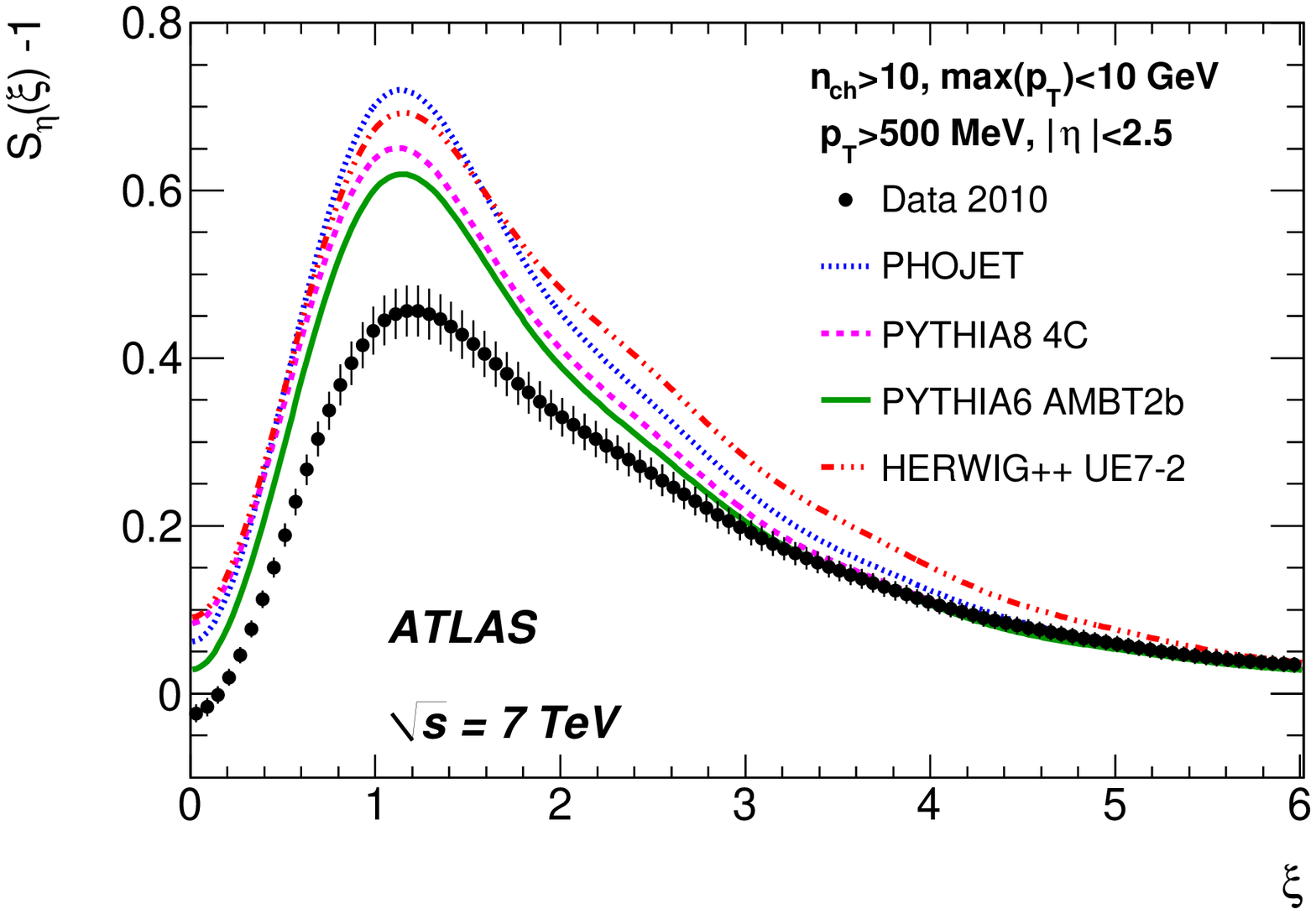}
 \caption{
   Corrected data from the low-p$_T$ depleted sample compared to particle-level predictions from various 
   MC models using conventional hadronisation algorithms.  The top and bottom plots are for the 
$S_{E}$ and $S_{\eta}$ power spectra, respectively. The error bars correspond to the 
combined statistical and systematic uncertainties.
\label{fig:corr_pt500}}
\end{center}
\end{figure}

Figure~\ref{fig:corr_pt500} shows the power spectra $S_{E}$ and $S_{\eta}$
for the corrected data and MC predictions in the low-$p_{T}$ depleted region 
($n_{ch}>10$, $p_{T}>$500 MeV and maximal $p_{T} <$~10~GeV).
In principle, this should be the best understood part of the phase space, 
with a suppressed diffractive component, lower sensitivity to hadronisation effects and best available model tunes. 
However, we find that all models significantly overestimate the size 
of the principal peak structure in both $S_{E}$ and $S_{\eta}$. 

Comparison of Figs 3, 4 and 5 show that the azimuthal correlations are qualitatively different in each subsample, and that the standard MC models fail to reproduce these accurately. 
Similar conclusion can be drawn for the measurement performed at $\sqrt{s}$~=~900~GeV (Appendix A).

\begin{figure}[ht!]
\begin{center}
\includegraphics[width=0.48\textwidth]{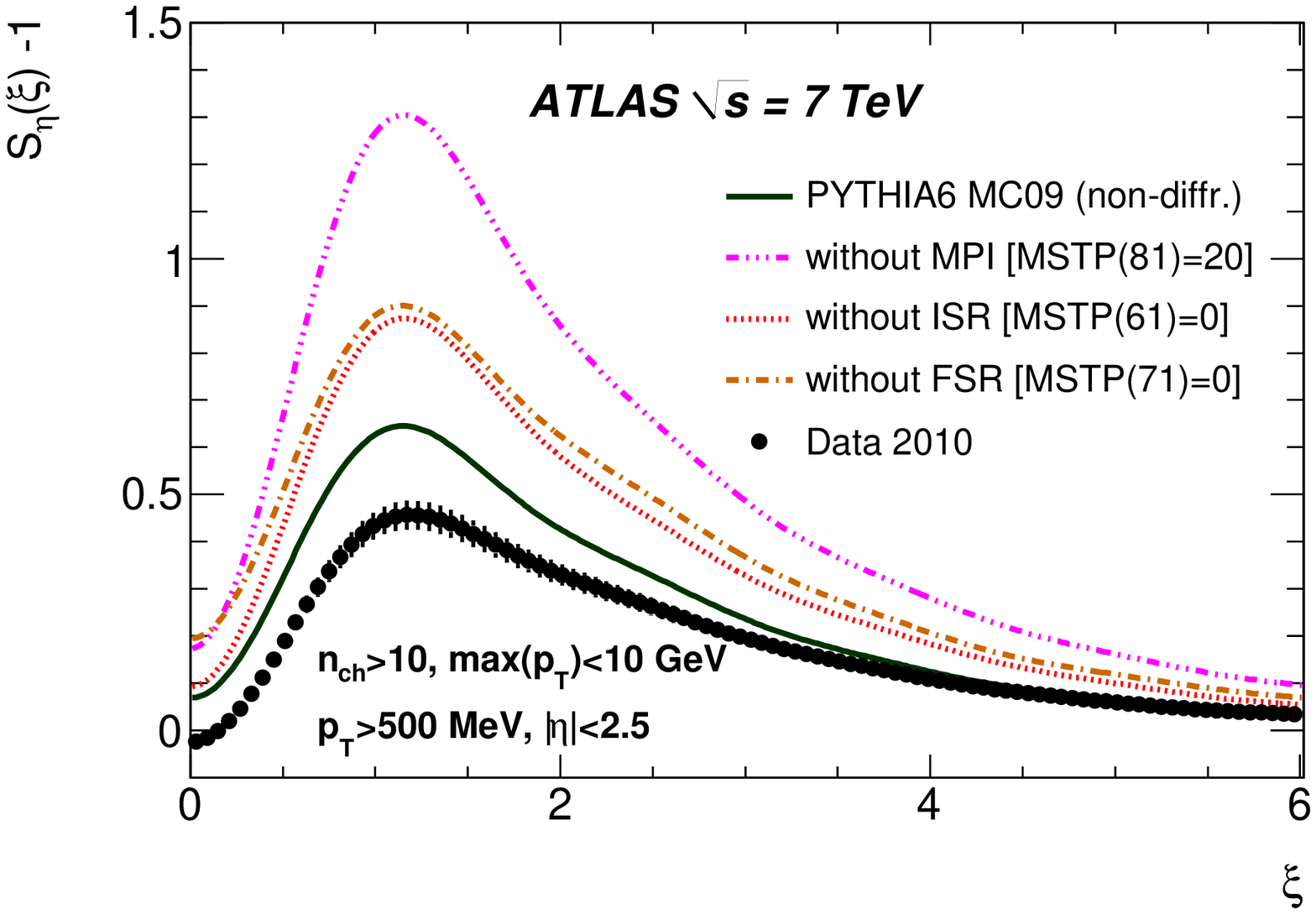}
\includegraphics[width=0.48\textwidth]{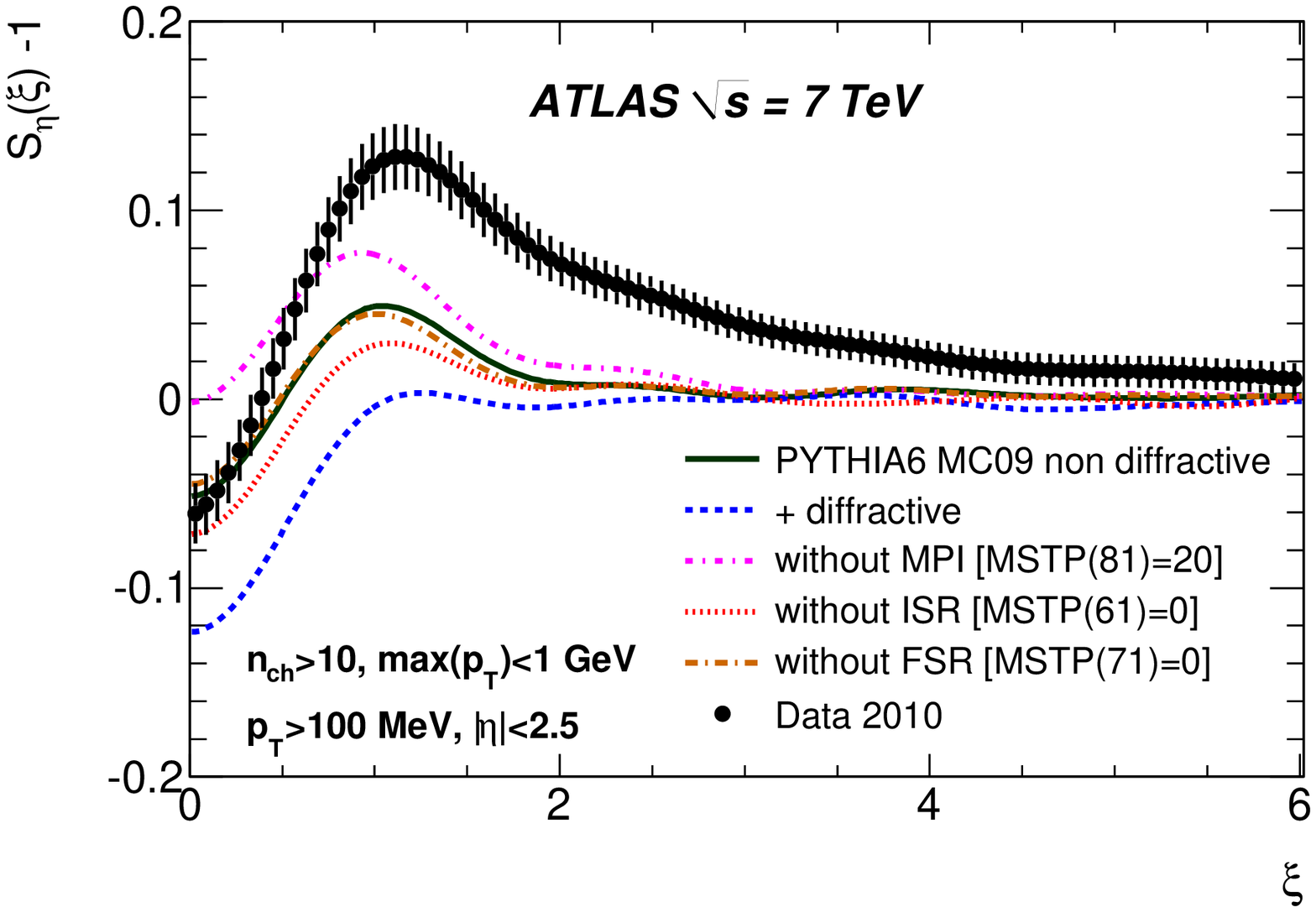}
\caption{
  Corrected $S_{\eta}$ distributions compared with the particle 
  level predictions of PYTHIA6 MC09 for various settings, using the non-diffractive {\it pp} scattering
 (full line) as the baseline. 
  Top: low-$p_T$ depleted subsample. Bottom: low-$p_T$ enhanced subsample.  
The error bars correspond to the combined statistical and systematic uncertainties. 
\label{fig:effects_comp}}
\end{center}
\end{figure}

  In the frame of the conventional QCD modelling, we have tried to identify the most likely source
of the observed discrepancies. Figure~\ref{fig:effects_comp} shows the sensitivity of the $S_{\eta}$ distribution
to various components of the QCD modelling implemented in PYTHIA 6, taking as a baseline the non-diffractive {\it pp} scattering scenario (indicated by the full line). In the low-p$_T$ depleted sample,
the size of correlations varies strongly with the amount of multiple parton interactions (MPI), of initial state radiation (ISR) and with the amount of parton showering. The data prefer modelling with enhanced radiation
and/or enhanced MPI rate which can be achieved via careful adjustment of the relevant model parameters.
   
   However, such an adjustment typically creates an even larger discrepancy in the low-p$_T$ enhanced region, where 
 the parton shower and ISR have smaller influence. Inversely, the removal of the MPI
 and of the diffractive processes increases the size of the peak in the modelling of the low-p$_T$ enhanced region,
 but none of these rather extreme variations lead to a good agreement with the measured $S_{\eta}$ distribution, while
 creating a huge discrepancy in the low-p$_T$ enhanced region. We conclude that for both measured spectra, the conventional
 models fail to describe the low-p$_T$ enhanced region, where the data are consistenly showing a larger and broader
 peak structure.

\subsection{Alternative fragmentation model}

The question of compatibility of the measurements with the azimuthal ordering signal 
originating from the underlying structure of the QCD field is studied using the PYTHIA6-based helix 
string model implementation \cite{pystrf} of the
modified helix string scenario corresponding to the $S_E$ definition (Eq.~2). 
\begin{figure}[b!]
\begin{center}
\includegraphics[width=0.48\textwidth]{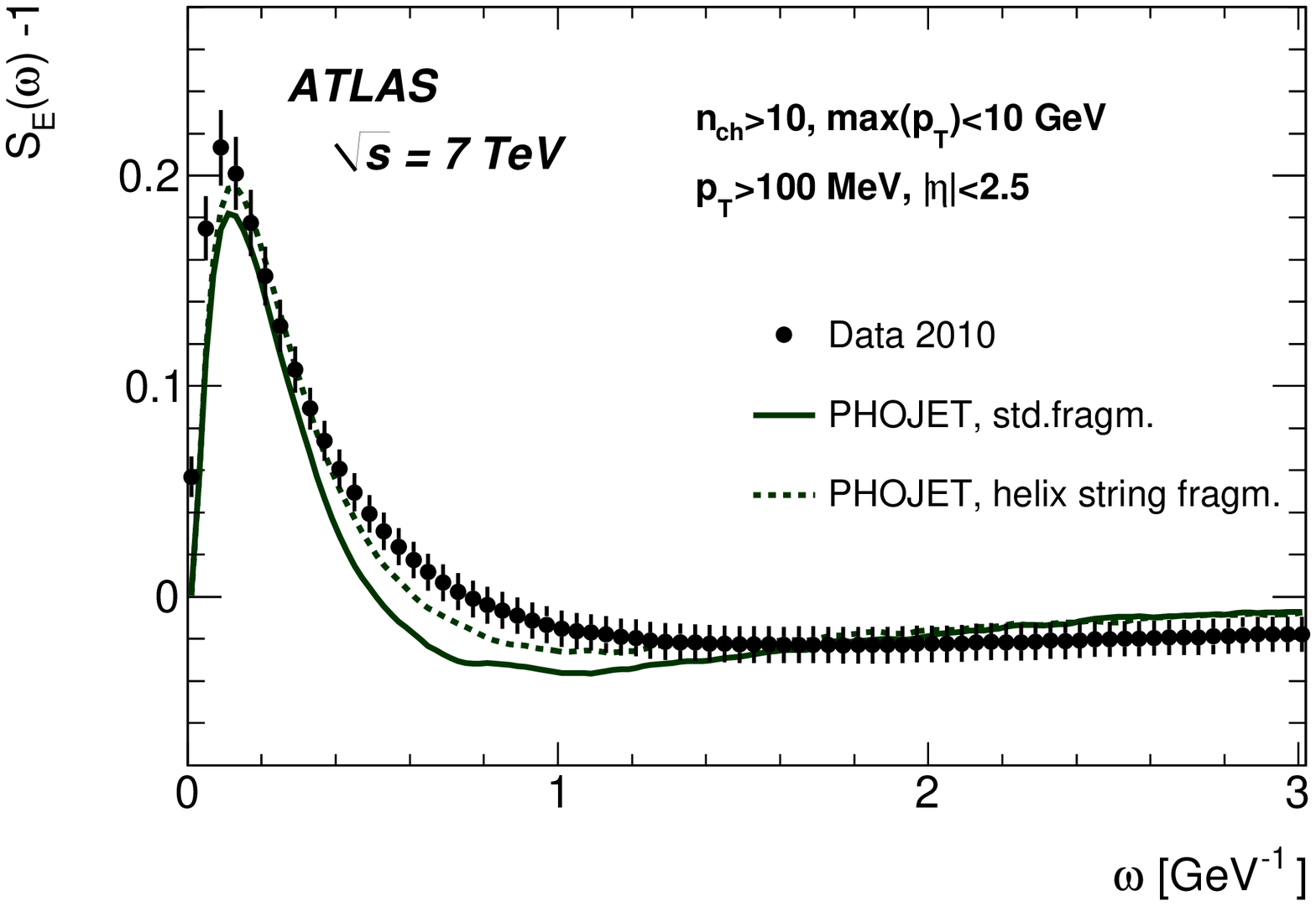}
\includegraphics[width=0.48\textwidth]{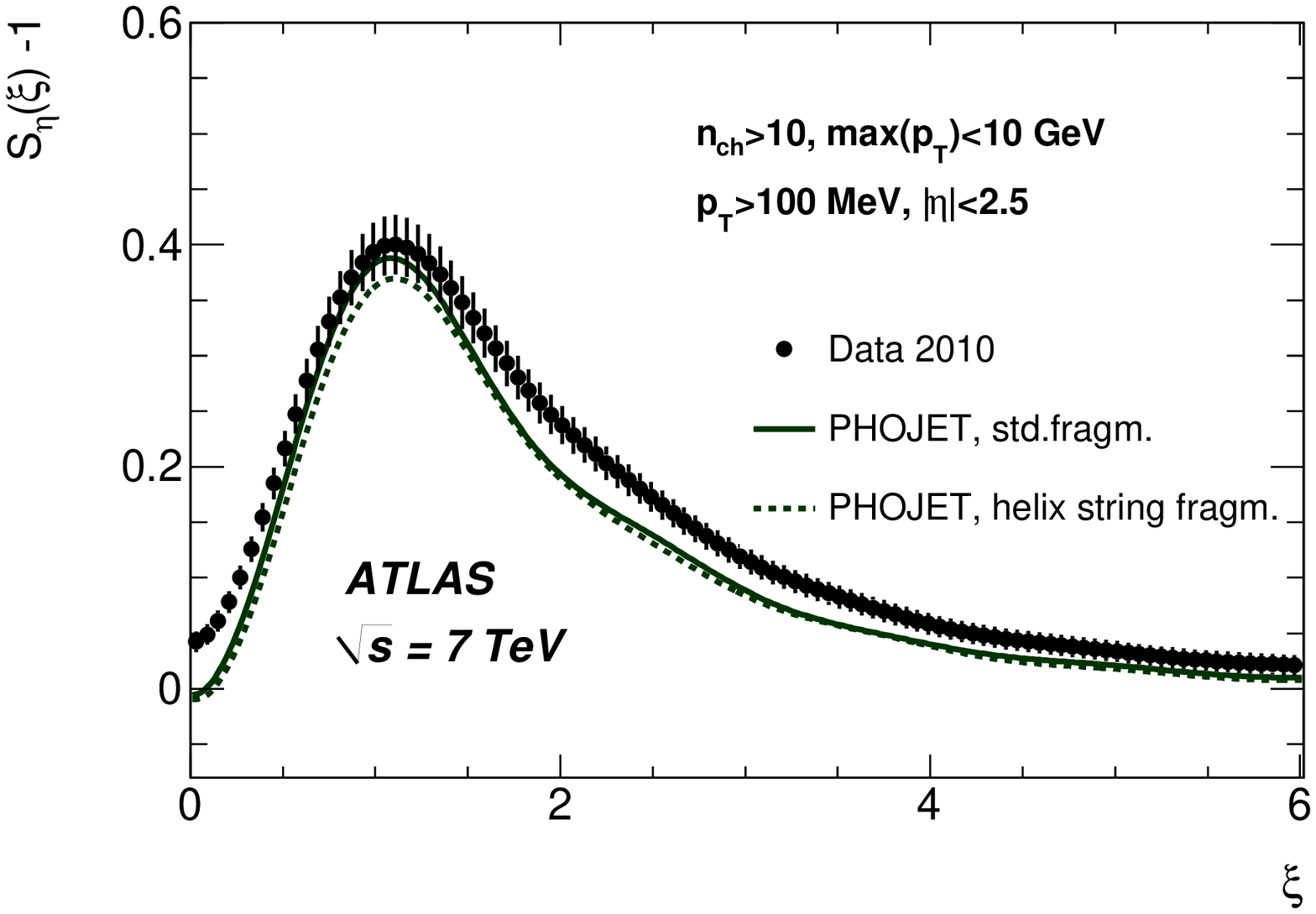}
\caption{
  Corrected $S_{E}$ (top) and $S_{\eta}$ (bottom) distributions compared with the particle 
  level predictions of PHOJET, interfaced to the PYTHIA6-based alternative fragmentation using helix string model ~\cite{pystrf}  (setting $\mathcal{S}$= 0.7 rad/GeV, see Eq.~2), in the inclusive event selection.
  The error bars correspond to the combined statistical and systematic uncertainties. 
\label{fig:unfolded_comp_incl}}
\end{center}
\end{figure}
The comparison of corrected data with the PHOJET modelling of {\it pp} interactions, interfaced alternatively
 with the standard string fragmentation and with the helix string fragmentation, is shown in 
Figs~\ref{fig:unfolded_comp_incl} and \ref{fig:unfolded_comp_lowpt}.
It is seen that the helix-like gluon ordering improves the description of the data in the inclusive sample
for $S_E$ as it generates higher values in the region of $\omega\in(0.5,1)$. In the low-p$_T$ enhanced sample
the data are more strongly peaked than the model. This indicates the data have a more jet-like structure
than predicted by PHOJET which means the model may need readjustment beyond the fragmentation part.    

It is possible that the original helix string proposal~\cite{lund_helixm} provides an improved description
of the $S_{\eta}$ measurement but we cannot verify this hypothesis due to the absence of the MC implementation
of this variant of the helix string model for parton configurations containing hard gluons.  

\begin{figure}[b!]
\begin{center}
\includegraphics[width=0.48\textwidth]{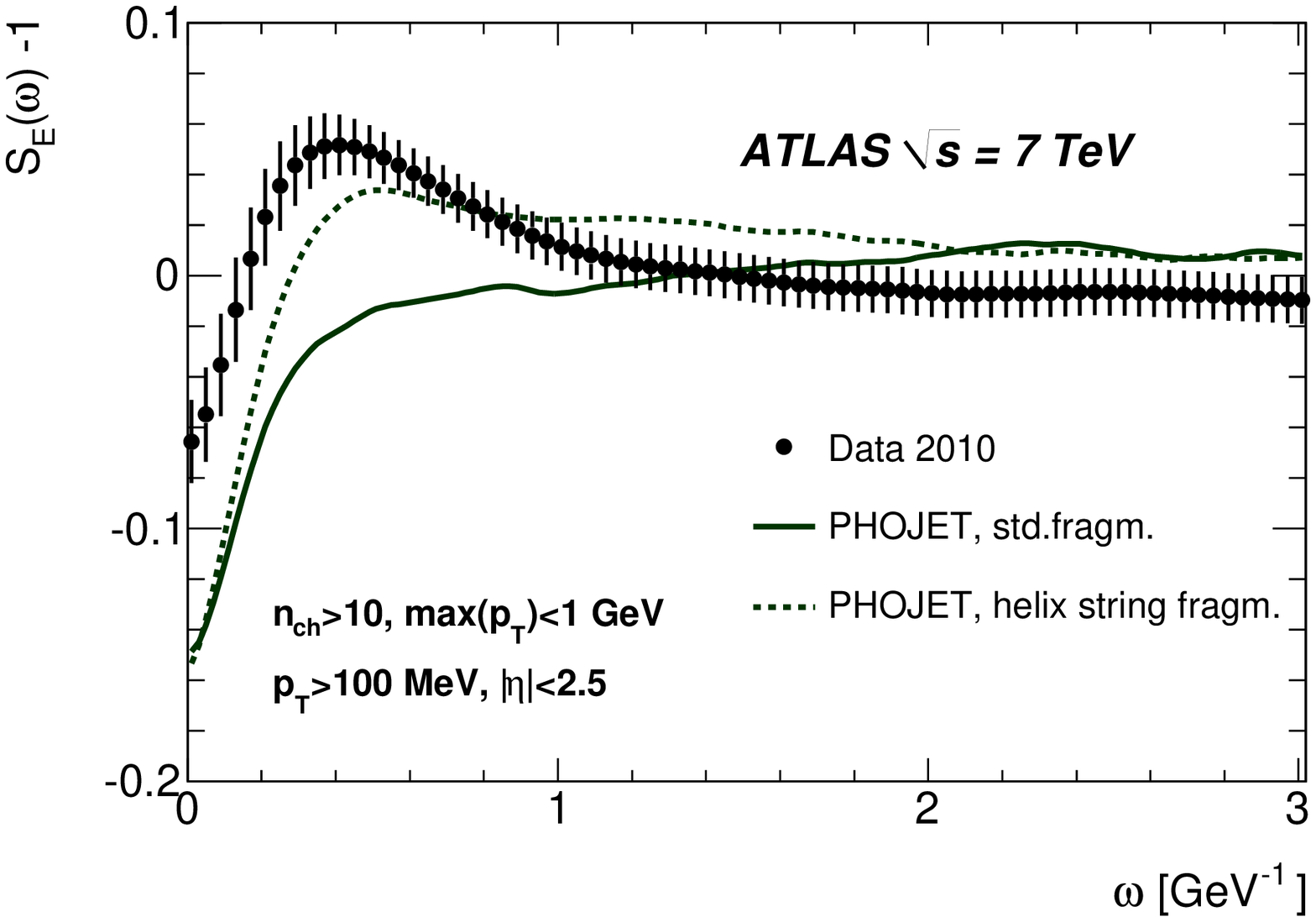}
\includegraphics[width=0.48\textwidth]{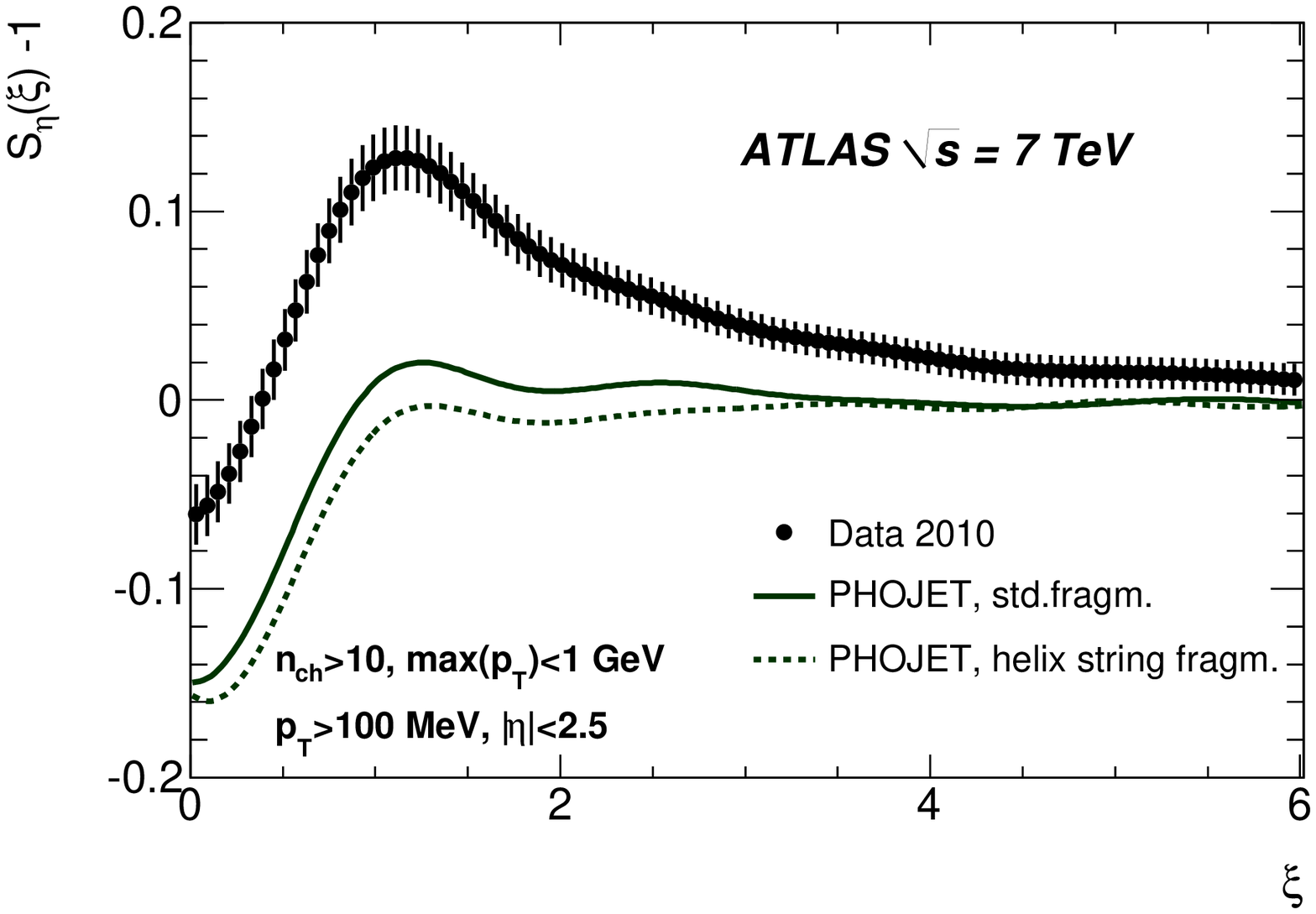}
\caption{
  Corrected $S_{E}$ (top) and $S_{\eta}$ (bottom) distributions compared with the particle 
  level predictions of PHOJET, interfaced to the PYTHIA6-based alternative fragmentation using helix string model ~\cite{pystrf}
  (setting $\mathcal{S}$= 0.7 rad/GeV, see Eq.~2), in the low-$p_{T}$ enhanced subsample. The error bars correspond to the combined statistical and systematic uncertainties. 
\label{fig:unfolded_comp_lowpt}}
\end{center}
\end{figure}

In principle, the position of the azimuthal ordering signal allows a measurement of the 
characteristic density of the helix string winding (parameter $\mathcal{S}$ in Eq.~2 ). The data 
indicate $\mathcal{S}$ in the range 0.5 to 0.9 rad/GeV. However, the interplay 
between string fragmentation and the other processes (such as parton showering, colour 
reconnection and multiple interactions) is not sufficiently understood to allow a more precise 
estimate (the measurement has to rely on MC description for the correction of the longitudinal boost).

%% file: conclusions.tex
\section {Conclusions}

A measurement of the ordering of charged hadrons in the azimuthal angle
with the ATLAS data recorded from proton-proton collisions at $\sqrt{s}$ = 7 TeV
and $\sqrt{s}$ = 900 GeV has been presented.

A spectral analysis of correlations between the opening azimuthal angle and the longitudinal separation
of the charged hadrons was performed by measuring the $S_{E}$ and $S_{\eta}$ power spectra.
These measurements were done in three kinematic regions (inclusive, low-$p_{T}$ enhanced 
and low-$p_{T}$ depleted samples) that were specifically defined to help assess the potential contribution of hadronisation effects to the power spectra by varying the levels of competition 
between hadronisation and other QCD effects.

The results were compared with the expectations of various MC event generators.   
Predictions generated by the MC models employing the standard Lund string fragmentation 
model roughly reproduce the data in the inclusive sample.

 The models systematically overestimate the size of correlations in the low-$p_{T}$ depleted sample, where the observables are sensitive to the multiple jet structure of events (due to the presence of underlying event and/or parton showering). 

For observables measured in the low-$p_{T}$ enhanced sample,
none of the models investigated describes the data adequately. 
A study showing the impact of extreme variations in MC model parameters 
that are known to contribute to soft-QCD effects demonstrates that, 
although some improvement of predictions for $S_{E}$ and $S_{\eta}$ can be achieved
in the low-$p_{T}$ enhanced sample, it is still far from satisfactory.
 
The measurement of $S_{E}$ in the kinematic region dominated by low-$p_{T}$ particles 
shows features similar to those seen in models in which the fragmenting QCD strings are 
represented by helix-like ordered gluon chains. These measurements suggest that the 
inclusion of such azimuthally-ordered fragmentation effects could be one factor in
improving current models of soft particle production and hadronisation.

%% file: acknowledgements.tex
\begin{acknowledgments}



We thank CERN for the very successful operation of the LHC, as well as the
support staff from our institutions without whom ATLAS could not be
operated efficiently.

We acknowledge the support of ANPCyT, Argentina; YerPhI, Armenia; ARC,
Australia; BMWF, Austria; ANAS, Azerbaijan; SSTC, Belarus; CNPq and FAPESP,
Brazil; NSERC, NRC and CFI, Canada; CERN; CONICYT, Chile; CAS, MOST and
NSFC, China; COLCIENCIAS, Colombia; MSMT CR, MPO CR and VSC CR, Czech
Republic; DNRF, DNSRC and Lundbeck Foundation, Denmark; ARTEMIS, European
Union; IN2P3-CNRS, CEA-DSM/IRFU, France; GNAS, Georgia; BMBF, DFG, HGF, MPG
and AvH Foundation, Germany; GSRT, Greece; ISF, MINERVA, GIF, DIP and
Benoziyo Center, Israel; INFN, Italy; MEXT and JSPS, Japan; CNRST, Morocco;
FOM and NWO, Netherlands; RCN, Norway; MNiSW, Poland; GRICES and FCT,
Portugal; MERYS (MECTS), Romania; MES of Russia and ROSATOM, Russian
Federation; JINR; MSTD, Serbia; MSSR, Slovakia; ARRS and MVZT, Slovenia;
DST/NRF, South Africa; MICINN, Spain; SRC and Wallenberg Foundation,
Sweden; SER, SNSF and Cantons of Bern and Geneva, Switzerland; NSC, Taiwan;
TAEK, Turkey; STFC, the Royal Society and Leverhulme Trust, United Kingdom;
DOE and NSF, United States of America.

The crucial computing support from all WLCG partners is acknowledged
gratefully, in particular from CERN and the ATLAS Tier-1 facilities at
TRIUMF (Canada), NDGF (Denmark, Norway, Sweden), CC-IN2P3 (France),
KIT/GridKA (Germany), INFN-CNAF (Italy), NL-T1 (Netherlands), PIC (Spain),
ASGC (Taiwan), RAL (UK) and BNL (USA) and in the Tier-2 facilities
worldwide.

\end{acknowledgments}

%% file: appendixA.tex
\appendix
\section*{Appendix A}

\begin{figure*}[b!]
\begin{center}
\includegraphics[width=0.49\textwidth]{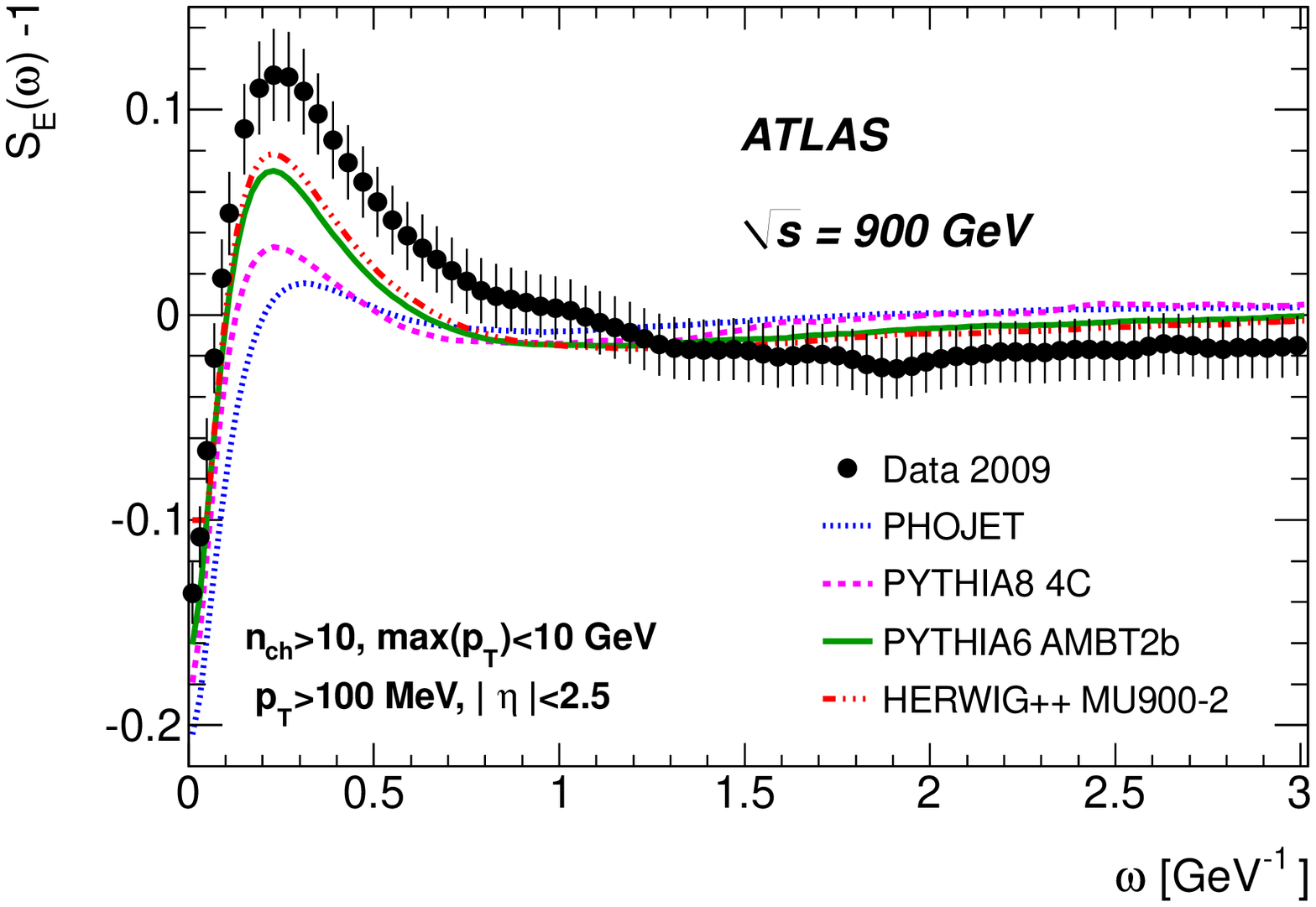}
\includegraphics[width=0.49\textwidth]{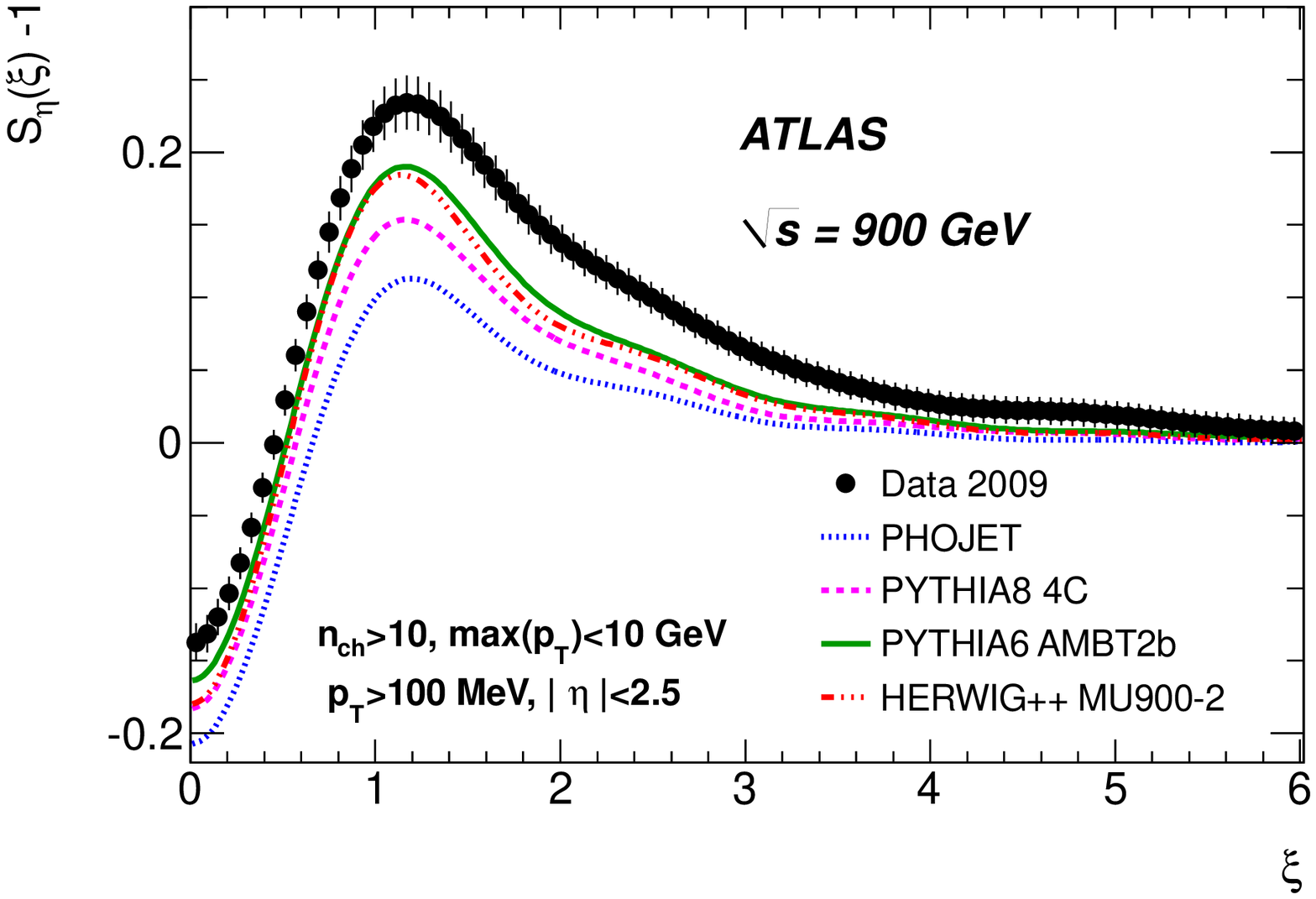}
\includegraphics[width=0.49\textwidth]{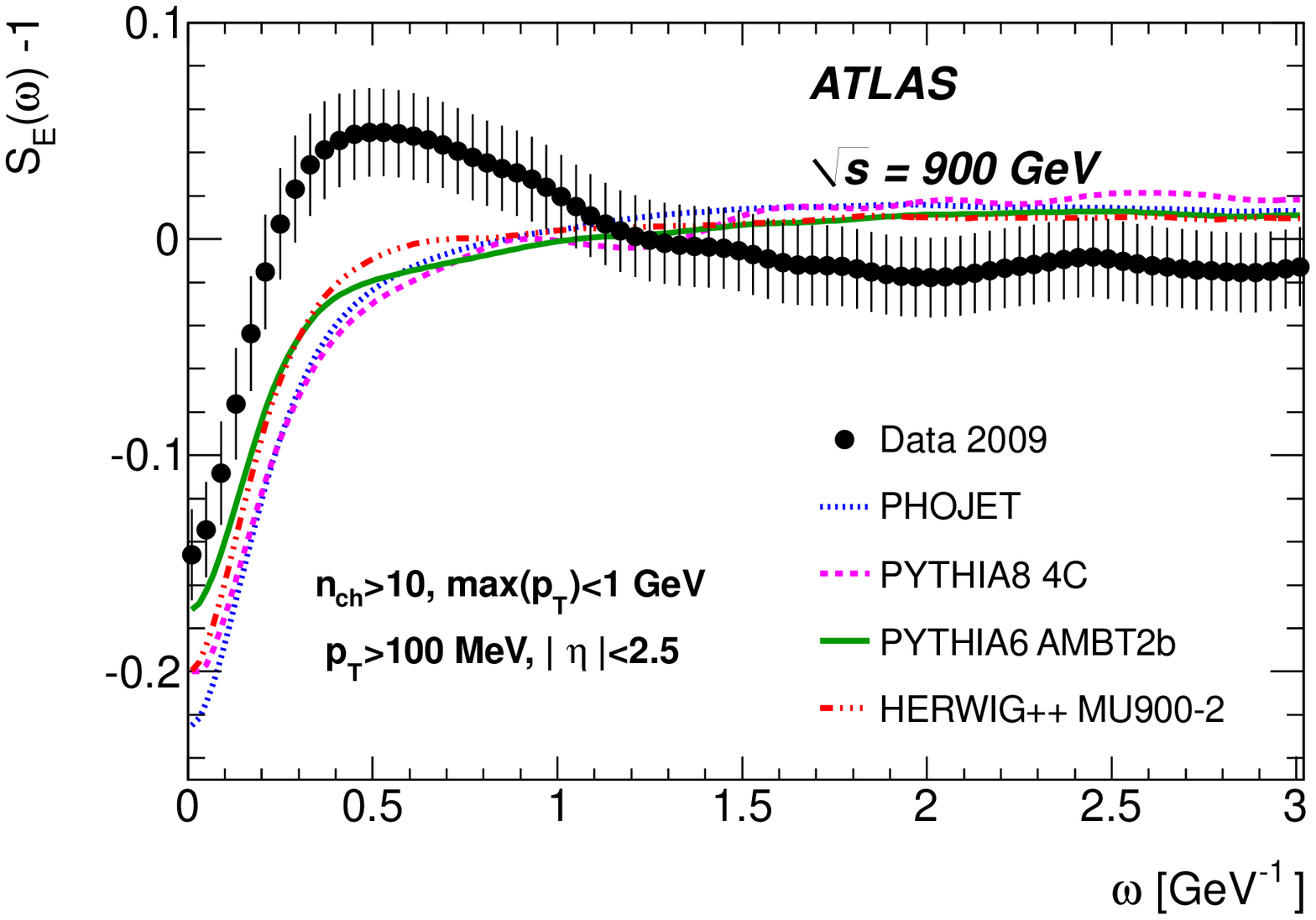}
\includegraphics[width=0.49\textwidth]{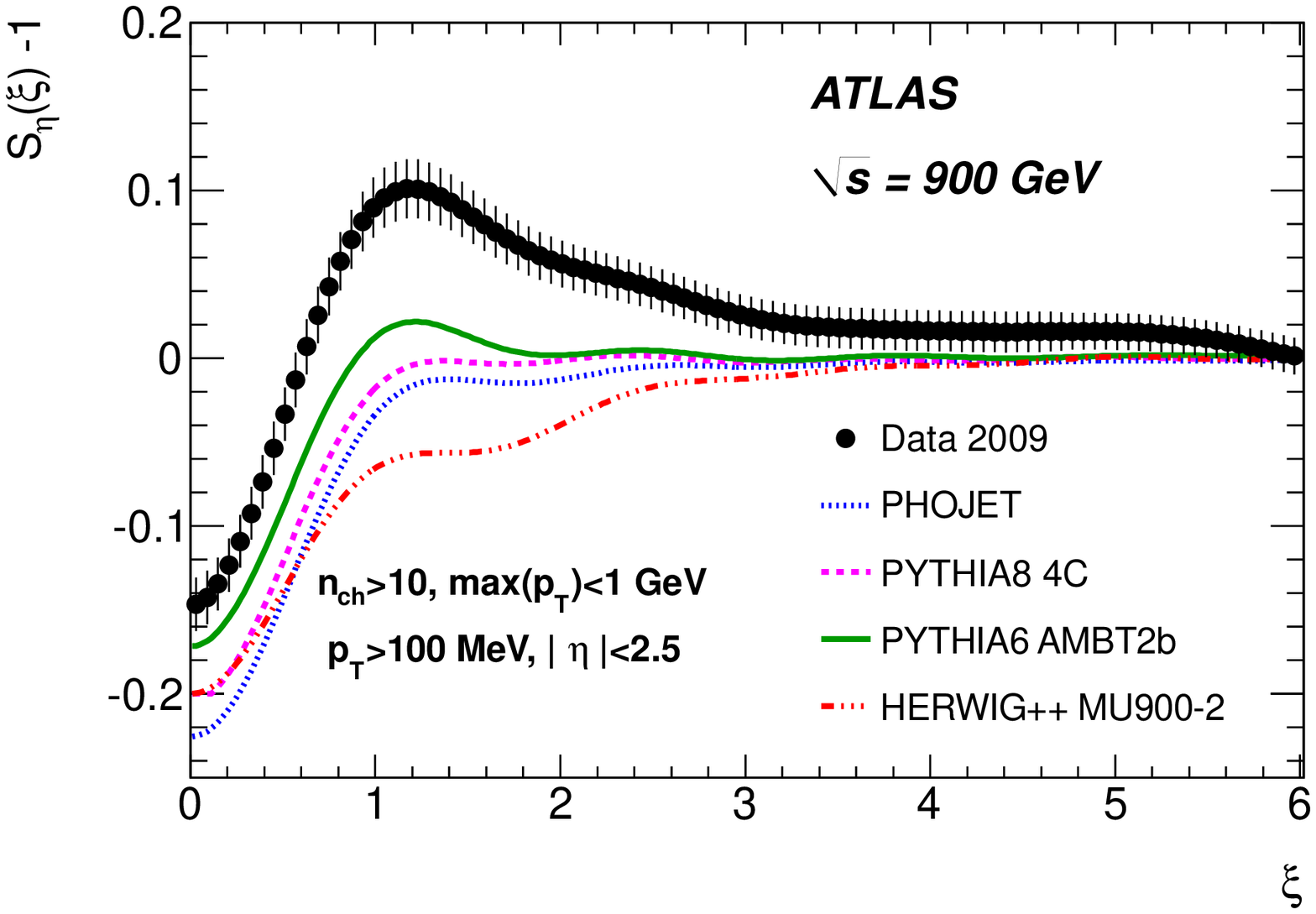}
\includegraphics[width=0.49\textwidth]{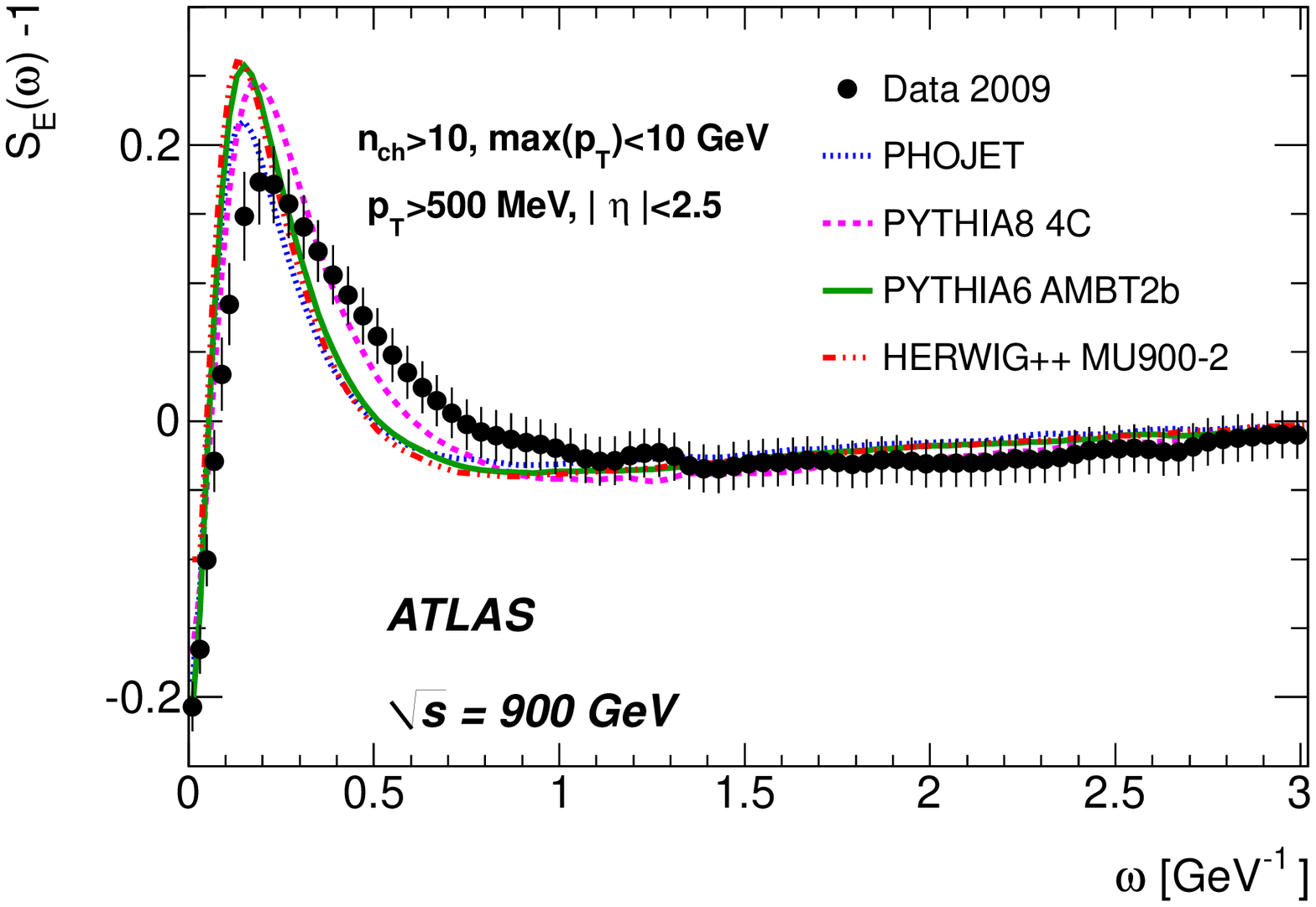}
\includegraphics[width=0.49\textwidth]{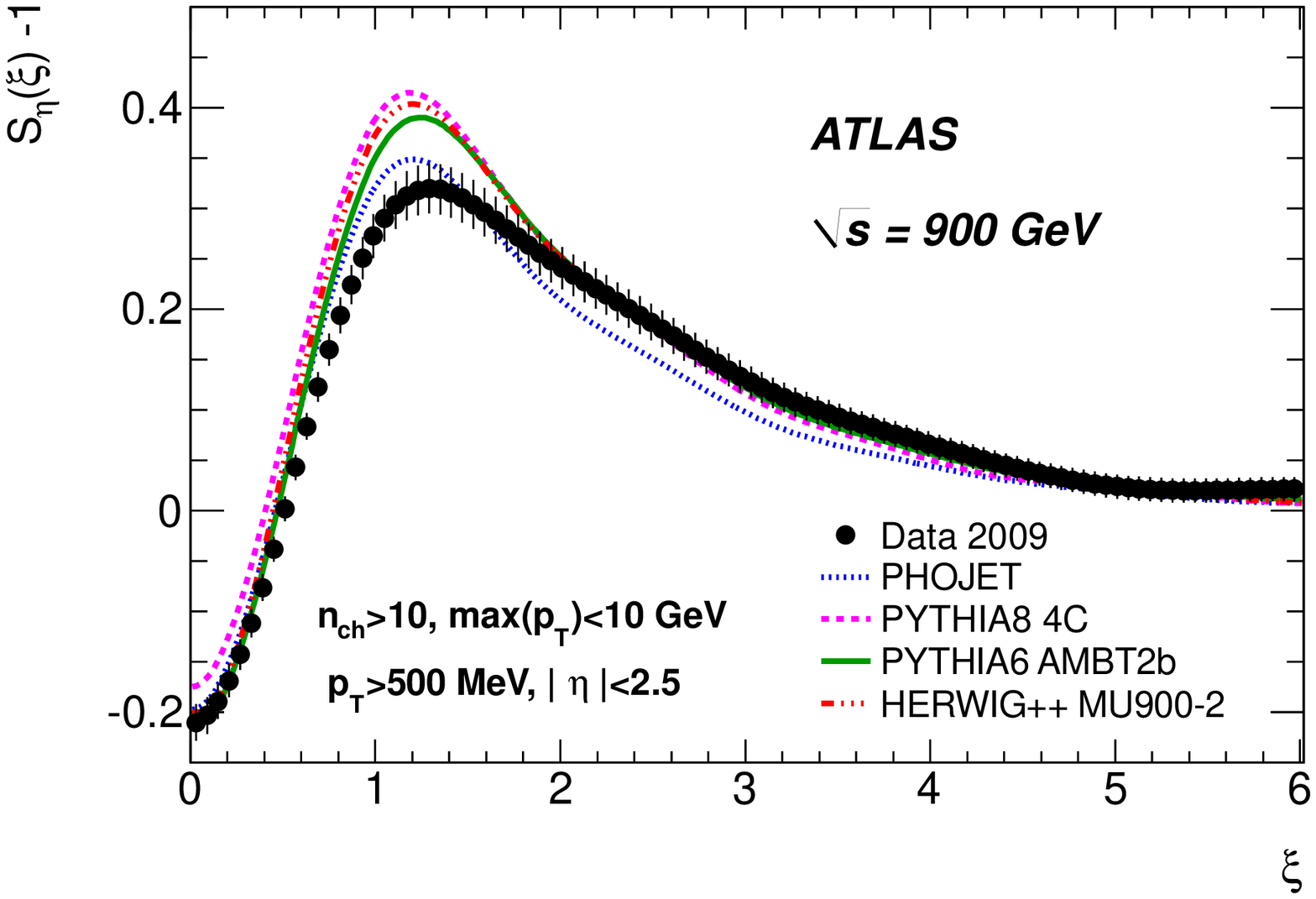}
 \caption{
   Corrected data compared to particle level predictions from various MC models using conventional hadronisation algorithms. The left and right plots are for the $S_{E}$ and $S_{\eta}$ power spectra, respectively. 
   Top: inclusive event selection ($p_T>$100 MeV, max($p_T)<$10 GeV).
   Middle: event selection with enhanced soft component ($p_T>$100 MeV, max($p_T)<$1 GeV).
   Bottom: event selection with reduced soft component ($p_T>$500 MeV, max($p_T)<$10 GeV).  
    ATLAS data collected at  $\sqrt{s}$ = 900 GeV and selected with $n_{ch}>10$. The error bars correspond 
to the combined statistical and systematic uncertainties.
\label{fig:data900}}
\end{center}
\end{figure*}

The analysis is repeated for ATLAS data collected at $\sqrt{s}=900$ GeV. The number 
of selected events and the average corrected charged-track multiplicity are shown in Table IV. 
For comparison, the average charged-particle multiplicity of PYTHIA6 MC09  samples is shown in Table V, 
together with the fraction of diffractive events in the sample (the estimate is based on the 
nominal cross section obtained from the generator). The corrected data are compared to the particle level prediction of various models in Fig.~\ref{fig:data900}.

\begin{table}[h!]{
\begin{center}
\begin{tabular}{|c|c|c|c|c|c|}
\hline
\multicolumn{6}{|c|} { \textit{pp} collisions at $\sqrt{s}$ = 900 GeV, \textit{n}$_{ch}>$10 } \\
\hline 
\multicolumn{4}{|c|} {  $p_T>$100 MeV  } &  \multicolumn{2}{|c|} { $p_T>$500 MeV }  \\
\hline 
\multicolumn{2}{|c|} {  max($p_T$)$<$10 GeV } &  \multicolumn{2}{|c|} { max($p_T$)$<$1 GeV } &   
\multicolumn{2}{|c|} { max($p_T$)$<$10 GeV  } \\
\hline
$ N_{ev} $   &  $ N_{ch} $   & $ N_{ev} $  &  $ N_{ch} $  &  $ N_{ev} $   &  $ N_{ch} $  \\
\hline
224717 & 24.8 & 59880  &  17.7$^*$ &  68456 & 16.3    \\        
\hline
\end{tabular}
\label{tab:sel_data900}
\caption{Number of selected events and average corrected charged-track multiplicity, per sample 
 ($^*$ indicates before the max($p_T$) cut correction).}
\end{center}}
\end{table}     
 
\begin{table}[h!]{
\begin{center}
\begin{tabular}{|l|c|c|c|}
\hline
 Model(tune) &  \multicolumn{3}{|c|}{ PYTHIA6(MC09)}  \\ 
\hline 
low $p_{T}$ cut &   \multicolumn{2}{|c|} {  $>$100 MeV  } &  $>$500 MeV \\
\hline
max($p_{T}$) cut &  $<$ 10 GeV & $<$ 1 GeV & $<$ 10 GeV \\   
\hline
 & \multicolumn{3}{|c|}{  n$_{ch}>$10  (corrected detector level)  } \\
\hline
$N^{gen}_{ch}$   &  21.91  &  15.28$^*$ & 15.2  \\
\hline
diffractive/total &  3.8($\pm$0.1)\%  &  14.7($\pm$0.2)\%  & $<$0.01\%  \\  
\hline
 & \multicolumn{3}{|c|}{  n$^{gen}_{ch}>$10  (particle level)  } \\
\hline
$N^{gen}_{ch}$   &  21.45  &  15.47 & 15.75  \\
\hline
diffractive/total &   3.7($\pm$0.1)\%  &  13.8($\pm$0.2)\%  & $<$0.01\%  \\  
\hline
\end{tabular}
\label{tab:sel_mc900}
\caption{Average charged-particle multiplicity $N^{gen}_{ch}$ and relative fraction of diffractive events
for the fully simulated PYTHIA6 (MC09) MC sample  at $\sqrt{s}=900$ GeV. 
Results are shown for events selected with the corrected charged-track multiplicity 
cut-off $n_{ch}>$10 (detector level) and  $n_{ch}^{gen}>$10 (particle level) ($^*$ indicates before the max($p_T$) cut correction). }
\end{center}}
\end{table}

%% file: appendixB.tex
\appendix
\section*{Appendix B} 

 The deconvolution technique employed in this analysis is a model-independent
 procedure suitable for observables with linear dependence
 on the track reconstruction efficiency.
 The dependence of the shape of the measured distribution on the fraction of reconstructed tracks
is studied by convolution with the track reconstruction efficiency matrix (``double-folding''): 
the reconstructed tracks are randomly rejected from the sample according to 
the parametrised reconstruction efficiency and the power spectra are recalculated. 
The procedure is repeated two more times, so that a sequence of three folding iterations 
is available for each measured distribution. The folding 
iterations, together with the measured distribution and the deconvoluted 
(MC truth) distribution, obey simple scaling rules:
\begin{itemize}
\item{ the size of correlations (along the S-1 axis) 
 scales linearly with the multiplicity 
 (physically it depends on the fraction of 
correlated pairs which have quadratic multiplicity dependence, partially 
compensated by the normalisation factor in Eqs.1 and 4). }
\item{ the shape of the power spectrum $S_E$ scales linearly in $\omega$ with the 
fraction of energy removed from the hadron chain (in $S_\eta$, the position of 
the peak stays nearly constant). }
\end{itemize}
\begin{figure}[b!]
\begin{center}
\includegraphics[width=0.45\textwidth]{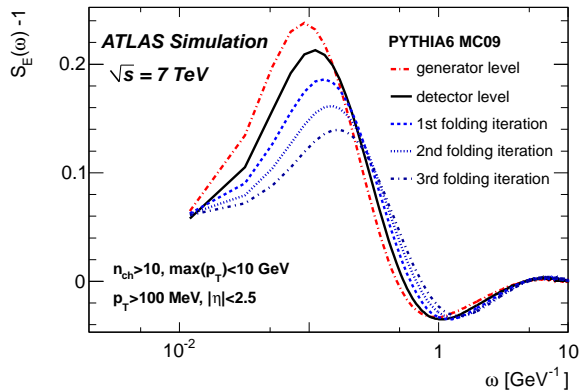}
 \caption{
Power spectrum calculated at the generator level, detector level and following
three folding iterations as described in the text. Secondary tracks
are removed from the samples in Figs 10-11. 
}
\label{fig:iter}
\end{center} 
\end{figure}
 A typical example of the variation of the power spectrum with the number of applied folding iterations is shown in
Fig.~\ref{fig:iter} on a PYTHIA6 sample with inclusive event selection. For better illustration of the scaling symmetry,
 secondary tracks are removed from the sample using the generator level information.  

 The deconvolution of the power spectrum requires finding a set of scaling factors $f_y, f_{x} (x=\omega,\xi)$ which fullfil the requirement
 \begin{equation}
    \tag{B1}
    S_{i-1}(x) -1 =  \frac{1}{f_{y}} [ S_{i}(\frac{x}{f_{x}}) -1 ] \ , \ 0 < x < 10 \ , \  0 < i < 4.
 \end{equation}
where $S_i$ designates the i-th folding iteration of the power spectrum
(i=0 for uncorrected data, -1 for corrected data). 

  The origin of observed correlations may be diverse, with variable dependence on the track reconstruction efficiency. A better precision can therefore be obtained by splitting the observed distribution empirically into components and by estimating the scaling factors per component.

  The power spectra calculated in this paper can, to good approximation,
 be fit with a combination of a Landau distribution 
\footnote{ $Landau(MPV,\sigma)=\frac{1}{\pi}\int_0^\infty e^{-u \text{ln} u - \lambda u} \  \text{sin} \ \pi ~ u ~\rm{d}u $,
where $\lambda=\frac{\omega-MPV}{\sigma}$. } (to describe the resonant peak structure), 
an exponentially falling background (driven by local momentum conservation of 
adjacent hadron pairs)  and a polynomial of first or second degree to 
describe the upper tail of the distribution. In practice, the subtraction of a single background term and determination of 2-3 effective scaling factors is sufficient to describe the difference between a pair of folding iterations; 
the replacement of the $f_x$ with an attenuated scaling factor  $f_x \rightarrow f^{'}_{x}=(f_{x}+x)/(1+x)$ 
helps smooth the transition into the upper part of the spectrum.    
\begin{figure}[b!]
\begin{center}
\includegraphics[width=0.45\textwidth]{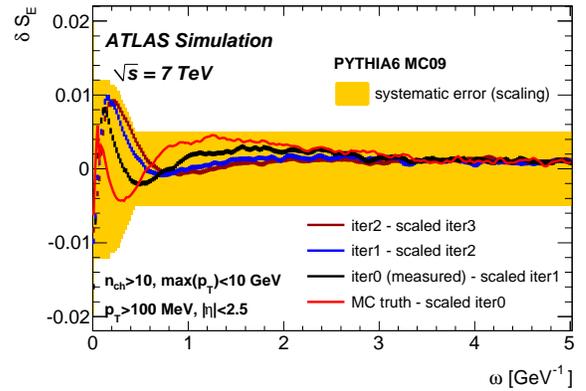}
 \caption{
The closure test of the scaling deconvolution method. Follow-up of 
the scaling precision over several folding iterations allows to verify the 
stability and to estimate the associated systematic error. 
\label{fig:systunf}}
\end{center} 
\end{figure}

 The knowledge of a single pair of consecutive 
iterations is sufficient to calculate the scaling factors. Typically,  they are determined using
the difference between uncorrected data and the 1st folding iteration (1$\rightarrow$0).
The stability of the scaling factors is verified by applying 
them on the folding iterations of higher orders (2$\rightarrow$1,3$\rightarrow$2) 
while the residual discrepancies serve as a basis for the estimate of the associated systematic 
uncertainty (Fig.~\ref{fig:systunf}). The deconvoluted power spectrum is obtained 
with the help of Eq.~B1 as the extrapolated (0$\rightarrow$-1) 
member of the sequence of folding iterations.

%% file: atlas_authlist_arXiv.tex
\begin{flushleft}
{\Large The ATLAS Collaboration}

\bigskip

G.~Aad$^{\rm 48}$,
B.~Abbott$^{\rm 110}$,
J.~Abdallah$^{\rm 11}$,
A.A.~Abdelalim$^{\rm 49}$,
A.~Abdesselam$^{\rm 117}$,
O.~Abdinov$^{\rm 10}$,
B.~Abi$^{\rm 111}$,
M.~Abolins$^{\rm 87}$,
O.S.~AbouZeid$^{\rm 157}$,
H.~Abramowicz$^{\rm 152}$
% ...
\bigskip

% ...
$^{10}$ Institute of Physics, Azerbaijan Academy of Sciences, Baku, Azerbaijan\\
$^{11}$ Institut de F\'isica d'Altes Energies and Departament de F\'isica de la Universitat Aut\`onoma  de Barcelona and ICREA, Barcelona, Spain\\
% ...
$^{48}$ Fakult\"{a}t f\"{u}r Mathematik und Physik, Albert-Ludwigs-Universit\"{a}t, Freiburg i.Br., Germany\\
$^{49}$ Section de Physique, Universit\'e de Gen\`eve, Geneva, Switzerland\\
% ...
$^{87}$ Department of Physics and Astronomy, Michigan State University, East Lansing MI, United States of America\\
% ...
$^{110}$ Homer L. Dodge Department of Physics and Astronomy, University of Oklahoma, Norman OK, United States of America\\
$^{111}$ Department of Physics, Oklahoma State University, Stillwater OK, United States of America\\
% ...
$^{117}$ Department of Physics, Oxford University, Oxford, United Kingdom\\
% ...
$^{152}$ Raymond and Beverly Sackler School of Physics and Astronomy, Tel Aviv University, Tel Aviv, Israel\\
% ...
$^{157}$ Department of Physics, University of Toronto, Toronto ON, Canada\\
% ...
\end{flushleft}


%% file: draft.bbl
\begin{thebibliography}{29}%
\makeatletter
\providecommand \@ifxundefined [1]{%
 \@ifx{#1\undefined}
}%
\providecommand \@ifnum [1]{%
 \ifnum #1\expandafter \@firstoftwo
 \else \expandafter \@secondoftwo
 \fi
}%
\providecommand \@ifx [1]{%
 \ifx #1\expandafter \@firstoftwo
 \else \expandafter \@secondoftwo
 \fi
}%
\providecommand \natexlab [1]{#1}%
\providecommand \enquote  [1]{``#1''}%
\providecommand \bibnamefont  [1]{#1}%
\providecommand \bibfnamefont [1]{#1}%
\providecommand \citenamefont [1]{#1}%
\providecommand \href@noop [0]{\@secondoftwo}%
\providecommand \href [0]{\begingroup \@sanitize@url \@href}%
\providecommand \@href[1]{\@@startlink{#1}\@@href}%
\providecommand \@@href[1]{\endgroup#1\@@endlink}%
\providecommand \@sanitize@url [0]{\catcode `\\12\catcode `\$12\catcode
  `\&12\catcode `\#12\catcode `\^12\catcode `\_12\catcode `\%12\relax}%
\providecommand \@@startlink[1]{}%
\providecommand \@@endlink[0]{}%
\providecommand \url  [0]{\begingroup\@sanitize@url \@url }%
\providecommand \@url [1]{\endgroup\@href {#1}{\urlprefix }}%
\providecommand \urlprefix  [0]{URL }%
\providecommand \Eprint [0]{\href }%
\providecommand \doibase [0]{http://dx.doi.org/}%
\providecommand \selectlanguage [0]{\@gobble}%
\providecommand \bibinfo  [0]{\@secondoftwo}%
\providecommand \bibfield  [0]{\@secondoftwo}%
\providecommand \translation [1]{[#1]}%
\providecommand \BibitemOpen [0]{}%
\providecommand \bibitemStop [0]{}%
\providecommand \bibitemNoStop [0]{.\EOS\space}%
\providecommand \EOS [0]{\spacefactor3000\relax}%
\providecommand \BibitemShut  [1]{\csname bibitem#1\endcsname}%
\let\auto@bib@innerbib\@empty
\bibitem [{\citenamefont {Collaboration}(2010{\natexlab{a}})}]{ATLAS-MB1}%
  \BibitemOpen
  \bibfield  {author} {\bibinfo {author} {\bibfnamefont {ATLAS}\ \bibnamefont
  {Collaboration}},\ }\href@noop {} {\bibfield  {journal} {\bibinfo  {journal}
  {Phys. Lett. B}\ }\textbf {\bibinfo {volume} {688}},\ \bibinfo {pages} {21}
  (\bibinfo {year} {2010}{\natexlab{a}})}\BibitemShut {NoStop}%
\bibitem [{\citenamefont {Collaboration}(2011{\natexlab{a}})}]{ATLAS-MB2}%
  \BibitemOpen
  \bibfield  {author} {\bibinfo {author} {\bibfnamefont {ATLAS}\ \bibnamefont
  {Collaboration}},\ }\href@noop {} {\bibfield  {journal} {\bibinfo  {journal}
  {New J. Phys.}\ }\textbf {\bibinfo {volume} {13}},\ \bibinfo {pages} {053033}
  (\bibinfo {year} {2011}{\natexlab{a}})}\BibitemShut {NoStop}%
\bibitem [{\citenamefont {Collaboration}(2010{\natexlab{b}})}]{ALICE-MB1}%
  \BibitemOpen
  \bibfield  {author} {\bibinfo {author} {\bibfnamefont {ALICE}\ \bibnamefont
  {Collaboration}},\ }\href@noop {} {\bibfield  {journal} {\bibinfo  {journal}
  {Eur. Phys. J. C}\ }\textbf {\bibinfo {volume} {65}},\ \bibinfo {pages} {111}
  (\bibinfo {year} {2010}{\natexlab{b}})}\BibitemShut {NoStop}%
\bibitem [{\citenamefont {Collaboration}(2010{\natexlab{c}})}]{ALICE-MB2}%
  \BibitemOpen
  \bibfield  {author} {\bibinfo {author} {\bibfnamefont {ALICE}\ \bibnamefont
  {Collaboration}},\ }\href@noop {} {\bibfield  {journal} {\bibinfo  {journal}
  {Eur. Phys. J. C}\ }\textbf {\bibinfo {volume} {68}},\ \bibinfo {pages} {89}
  (\bibinfo {year} {2010}{\natexlab{c}})}\BibitemShut {NoStop}%
\bibitem [{\citenamefont {Collaboration}(2010{\natexlab{d}})}]{ALICE-MB3}%
  \BibitemOpen
  \bibfield  {author} {\bibinfo {author} {\bibfnamefont {ALICE}\ \bibnamefont
  {Collaboration}},\ }\href@noop {} {\bibfield  {journal} {\bibinfo  {journal}
  {Eur. Phys. J. C}\ }\textbf {\bibinfo {volume} {68}},\ \bibinfo {pages} {345}
  (\bibinfo {year} {2010}{\natexlab{d}})}\BibitemShut {NoStop}%
\bibitem [{\citenamefont {Collaboration}(2010{\natexlab{e}})}]{CMS-MB1}%
  \BibitemOpen
  \bibfield  {author} {\bibinfo {author} {\bibfnamefont {CMS}\ \bibnamefont
  {Collaboration}},\ }\href@noop {} {\bibfield  {journal} {\bibinfo  {journal}
  {J. High Energy Phys.}\ }\textbf {\bibinfo {volume} {02}},\ \bibinfo {pages}
  {041} (\bibinfo {year} {2010}{\natexlab{e}})}\BibitemShut {NoStop}%
\bibitem [{\citenamefont {Collaboration}(2010{\natexlab{f}})}]{CMS-MB2}%
  \BibitemOpen
  \bibfield  {author} {\bibinfo {author} {\bibfnamefont {CMS}\ \bibnamefont
  {Collaboration}},\ }\href@noop {} {\bibfield  {journal} {\bibinfo  {journal}
  {Phys. Rev. Lett.}\ }\textbf {\bibinfo {volume} {105}},\ \bibinfo {pages}
  {022002} (\bibinfo {year} {2010}{\natexlab{f}})}\BibitemShut {NoStop}%
\bibitem [{\citenamefont {Collaboration}(2011{\natexlab{b}})}]{CMS-MB3}%
  \BibitemOpen
  \bibfield  {author} {\bibinfo {author} {\bibfnamefont {CMS}\ \bibnamefont
  {Collaboration}},\ }\href@noop {} {\bibfield  {journal} {\bibinfo  {journal}
  {J. High Energy Phys.}\ }\textbf {\bibinfo {volume} {01}},\ \bibinfo {pages}
  {079} (\bibinfo {year} {2011}{\natexlab{b}})}\BibitemShut {NoStop}%
\bibitem [{\citenamefont {\textit{et al.}}(1985)}]{duality}%
  \BibitemOpen
  \bibfield  {author} {\bibinfo {author} {\bibfnamefont {Y.~I.~Azimov}\
  \bibnamefont {\textit{et al.}}},\ }\href@noop {} {\bibfield  {journal}
  {\bibinfo  {journal} {Z. Phys. C}\ }\textbf {\bibinfo {volume} {27}},\
  \bibinfo {pages} {65} (\bibinfo {year} {1985})}\BibitemShut {NoStop}%
\bibitem [{\citenamefont {\textit{et al.}}(1983)}]{lund}%
  \BibitemOpen
  \bibfield  {author} {\bibinfo {author} {\bibfnamefont {B.~Andersson}\
  \bibnamefont {\textit{et al.}}},\ }\href@noop {} {\bibfield  {journal}
  {\bibinfo  {journal} {Phys. Rept.}\ }\textbf {\bibinfo {volume} {97}},\
  \bibinfo {pages} {31} (\bibinfo {year} {1983})}\BibitemShut {NoStop}%
\bibitem [{\citenamefont {Webber}(1984)}]{cluster01}%
  \BibitemOpen
  \bibfield  {author} {\bibinfo {author} {\bibfnamefont {B.~R.}\ \bibnamefont
  {Webber}},\ }\href@noop {} {\bibfield  {journal} {\bibinfo  {journal} {Nucl.
  Phys. B}\ }\textbf {\bibinfo {volume} {238}},\ \bibinfo {pages} {492}
  (\bibinfo {year} {1984})}\BibitemShut {NoStop}%
\bibitem [{\citenamefont {\textit{et al.}}(1998)}]{lund_helixm}%
  \BibitemOpen
  \bibfield  {author} {\bibinfo {author} {\bibfnamefont {B.~Andersson}\
  \bibnamefont {\textit{et al.}}},\ }\href@noop {} {\bibfield  {journal}
  {\bibinfo  {journal} {J. High Energy Phys.}\ }\textbf {\bibinfo {volume}
  {09}},\ \bibinfo {pages} {014} (\bibinfo {year} {1998})}\BibitemShut
  {NoStop}%
\bibitem [{Note1()}]{Note1}%
  \BibitemOpen
  \bibinfo {note} {The ATLAS reference system is a right-handed coordinate
  system with its origin at the nominal interaction point at the centre of the
  detector. Cylindrical coordinates ($r,\phi $) are used in the transverse
  plane, $\phi $ being the azimuthal angle around the beam axis. The
  pseudorapidity is defined in terms of the polar angle $\theta $ as $\eta =
  -\protect \qopname \relax o{ln}\protect \qopname \relax o{tan}~( \theta /2
  )$.}\BibitemShut {Stop}%
\bibitem [{\citenamefont {\v{S}. Todorova}()}]{ismd}%
  \BibitemOpen
  \bibfield  {author} {\bibinfo {author} {\bibnamefont {\v{S}. Todorova}},\
  }\href@noop {} {}\bibinfo {note} {$\text{in}$ \textit{ Proceedings of the XL
  International Symposium on Multiparticle Dynamics, Antwerp, 2010,} edited by
  P.van Mechelen and N.van Remortel, published by Uitgeverij UPA University
  Press Antwerp (2011)}\BibitemShut {NoStop}%
\bibitem [{\citenamefont {Collaboration}(2008)}]{ATLAS}%
  \BibitemOpen
  \bibfield  {author} {\bibinfo {author} {\bibfnamefont {ATLAS}\ \bibnamefont
  {Collaboration}},\ }\href@noop {} {\bibfield  {journal} {\bibinfo  {journal}
  {JINST}\ }\textbf {\bibinfo {volume} {3}},\ \bibinfo {pages} {S08003}
  (\bibinfo {year} {2008})}\BibitemShut {NoStop}%
\bibitem [{\citenamefont {Sj$\ddot{\mathrm{o}}$strand}\ \emph
  {et~al.}(2006)\citenamefont {Sj$\ddot{\mathrm{o}}$strand}, \citenamefont
  {Mrenna},\ and\ \citenamefont {Skands}}]{PYTHIA}%
  \BibitemOpen
  \bibfield  {author} {\bibinfo {author} {\bibfnamefont {T.}~\bibnamefont
  {Sj$\ddot{\mathrm{o}}$strand}}, \bibinfo {author} {\bibfnamefont
  {S.}~\bibnamefont {Mrenna}}, \ and\ \bibinfo {author} {\bibfnamefont {P.~Z.}\
  \bibnamefont {Skands}},\ }\href@noop {} {\bibfield  {journal} {\bibinfo
  {journal} {J. High Energy Phys.}\ }\textbf {\bibinfo {volume} {05}},\
  \bibinfo {pages} {026} (\bibinfo {year} {2006})}\BibitemShut {NoStop}%
\bibitem [{\citenamefont {Collaboration}(2010{\natexlab{g}})}]{ATLAS-MC09}%
  \BibitemOpen
  \bibfield  {author} {\bibinfo {author} {\bibfnamefont {ATLAS}\ \bibnamefont
  {Collaboration}},\ }\href@noop {} {} (\bibinfo {year} {2010}{\natexlab{g}}),\
  \bibinfo {note} {$\text{Report No. ATL-PHYS-PUB-2010-002}$,
  $\text{https://cdsweb.cern.ch/record/1247375.}$}\BibitemShut {NoStop}%
\bibitem [{\citenamefont {Collaboration}(2003)}]{GEANT4}%
  \BibitemOpen
  \bibfield  {author} {\bibinfo {author} {\bibfnamefont {GEANT4}\ \bibnamefont
  {Collaboration}},\ }\href@noop {} {\bibfield  {journal} {\bibinfo  {journal}
  {Nucl. Instr. Meth.}\ }\textbf {\bibinfo {volume} {A506}},\ \bibinfo {pages}
  {250} (\bibinfo {year} {2003})}\BibitemShut {NoStop}%
\bibitem [{\citenamefont {Collaboration}(2010{\natexlab{h}})}]{ATLAS-simul}%
  \BibitemOpen
  \bibfield  {author} {\bibinfo {author} {\bibfnamefont {ATLAS}\ \bibnamefont
  {Collaboration}},\ }\href@noop {} {\bibfield  {journal} {\bibinfo  {journal}
  {Eur. Phys. J. C}\ }\textbf {\bibinfo {volume} {70}},\ \bibinfo {pages} {823}
  (\bibinfo {year} {2010}{\natexlab{h}})}\BibitemShut {NoStop}%
\bibitem [{\citenamefont {Engel}(1995)}]{PHOJET}%
  \BibitemOpen
  \bibfield  {author} {\bibinfo {author} {\bibfnamefont {R.}~\bibnamefont
  {Engel}},\ }\href@noop {} {\bibfield  {journal} {\bibinfo  {journal} {Z.
  Phys. C}\ }\textbf {\bibinfo {volume} {66}},\ \bibinfo {pages} {203}
  (\bibinfo {year} {1995})}\BibitemShut {NoStop}%
\bibitem [{\citenamefont {\textit{et al.}}(2008)}]{HERWIG++}%
  \BibitemOpen
  \bibfield  {author} {\bibinfo {author} {\bibfnamefont {M.~Bahr}\ \bibnamefont
  {\textit{et al.}}},\ }\href@noop {} {\bibfield  {journal} {\bibinfo
  {journal} {Eur. Phys. J. C}\ }\textbf {\bibinfo {volume} {58}},\ \bibinfo
  {pages} {639} (\bibinfo {year} {2008})}\BibitemShut {NoStop}%
\bibitem [{hwg()}]{hwgtune}%
  \BibitemOpen
  \href@noop {} {}\bibinfo {note}
  {$\text{http://projects.hepforge.org/herwig/trac/wiki/MB\_UE\_tunes}$}\BibitemShut
  {NoStop}%
\bibitem [{\citenamefont {Sj$\ddot{\mathrm{o}}$strand}\ \emph
  {et~al.}(2008)\citenamefont {Sj$\ddot{\mathrm{o}}$strand}, \citenamefont
  {Mrenna},\ and\ \citenamefont {Skands}}]{PYTHIA8}%
  \BibitemOpen
  \bibfield  {author} {\bibinfo {author} {\bibfnamefont {T.}~\bibnamefont
  {Sj$\ddot{\mathrm{o}}$strand}}, \bibinfo {author} {\bibfnamefont
  {S.}~\bibnamefont {Mrenna}}, \ and\ \bibinfo {author} {\bibfnamefont {P.~Z.}\
  \bibnamefont {Skands}},\ }\href@noop {} {\bibfield  {journal} {\bibinfo
  {journal} {Comput. Phys. Comm.}\ }\textbf {\bibinfo {volume} {178}},\
  \bibinfo {pages} {852} (\bibinfo {year} {2008})}\BibitemShut {NoStop}%
\bibitem [{\citenamefont {Collaboration}()}]{AMBT2b}%
  \BibitemOpen
  \bibfield  {author} {\bibinfo {author} {\bibfnamefont {ATLAS}\ \bibnamefont
  {Collaboration}},\ }\href@noop {} {}\bibinfo {note}
  {$\text{ATL-PHYS-PUB-2011-009}$, $\text{(2011)}$,
  $\text{https://cdsweb.cern.ch/record/1363300}$}\BibitemShut {NoStop}%
\bibitem [{pys()}]{pystrf}%
  \BibitemOpen
  \href@noop {} {}\bibinfo {note}
  {$\text{http://projects.hepforge.org/helix/}$}\BibitemShut {NoStop}%
\bibitem [{Note2()}]{Note2}%
  \BibitemOpen
  \bibinfo {note} {Primary particles are defined as all particles with lifetime
  longer than 0.3$\times $10$^{-10}$s originating from the primary interaction
  or from subsequent decay of particles with shorter lifetime.}\BibitemShut
  {Stop}%
\bibitem [{Note3()}]{Note3}%
  \BibitemOpen
  \bibinfo {note} {The track reconstruction efficiency is parametrised as a
  function of track $p_{T}$ and pseudorapidity $\eta $}\BibitemShut {NoStop}%
\bibitem [{unf()}]{unfolding}%
  \BibitemOpen
  \href@noop {} {}\bibinfo {note} {$\text{J.~W.~Monk and C.~Oropeza-Barrera,
  arXiv:1111.4896v2}$ $\text{[NIM (to be published)]}$}\BibitemShut {NoStop}%
\bibitem [{Note4()}]{Note4}%
  \BibitemOpen
  \bibinfo {note} {$Landau(MPV,\sigma )=\protect \frac {1}{\pi }\DOTSI \intop
  \ilimits@ _0^\infty e^{-u \protect \text {ln} u - \lambda u} \ \protect \text
  {sin} \ \pi ~ u ~\protect \rm {d}u $, where $\lambda =\protect \frac {\omega
  -MPV}{\sigma }$.}\BibitemShut {Stop}%
\end{thebibliography}%
